# The Impact of Privacy Laws on Online User Behavior

**Julia Schmitt, Klaus M. Miller, Bernd Skiera**
**Goethe University Frankfurt**

Julia Schmitt, Department of Marketing, Faculty of Economics and Business, Goethe University Frankfurt, Theodor-W.-Adorno-Platz 4, 60323 Frankfurt, Germany, Phone +49-69-798-34563, Email: schmitt@wiwi.uni-frankfurt.de.

Prof. Dr. Klaus M. Miller, Department of Marketing, HEC Paris, 1 Rue de la Libération, 78350 Jouy-en-Josas, France, Phone +33-139-677-088, Email: millerk@hec.fr.

Prof. Dr. Bernd Skiera, Department of Marketing, Faculty of Economics and Business, Goethe University Frankfurt, Theodor-W.-Adorno-Platz 4, 60323 Frankfurt, Germany, Phone +49-69-798-34649, Email: skiera@wiwi.uni-frankfurt.de.

# The Impact of Privacy Laws on Online User Behavior

*ABSTRACT*

Policymakers worldwide draft privacy laws that require trading-off between safeguarding consumer privacy and preventing economic loss to companies that use consumer data. However, little empirical knowledge exists as to how privacy laws affect companies' performance. Accordingly, this paper empirically quantifies the effects of the enforcement of the EU's General Data Protection Regulation (GDPR) on online user behavior over time, analyzing data from 6,286 websites spanning 24 industries during the 10 months before and 18 months after the GDPR's enforcement in 2018. A panel differences estimator, with a synthetic control group approach, isolates the short- and long-term effects of the GDPR on user behavior. The results show that, on average, the GDPR's effects on user quantity and usage intensity are negative; e.g., the numbers of total visits to a website decrease by 4.9% and 10% due to GDPR in respectively the short- and long-term. These effects could translate into average revenue losses of $7 million for e-commerce websites and almost $2.5 million for ad-based websites 18 months after GDPR. The GDPR's effects vary across websites, with some industries even benefiting from it; moreover, more-popular websites suffer less, suggesting that the GDPR increased market concentration.

Internet users generally perceive their privacy as a cause for concern. For example, a survey in 2019 by Pew Research Center showed that 79% of American users are concerned about how companies use their data, partly because they do not know which data companies collect. In recent years, policymakers worldwide have drafted and enforced privacy laws to mitigate these types of concerns.

One of the highest-profile and most expansive laws is the European Union's (EU) General Data Protection Regulation (GDPR), which became enforceable on May 25th, 2018. Similarly, other countries such as Chile, Serbia, Brazil, India, and Thailand have also recently enforced or approved privacy laws. While the specific details of the various privacy laws differ, their basic idea is to increase the individuals' privacy, commonly defined as the individuals' control over their personal data (Holvast 1993).

In practical terms, privacy laws such as the GDPR seek to enhance data privacy by targeting the operations of companies that handle user data through two main avenues: 1) limiting companies' capacity to collect and use user data, and 2) requiring that companies be transparent about their data collection practices.

On the one hand, these requirements resulted in websites reducing the number of third-party cookies (e.g., Libert et al. 2018) and updating and providing more information in their privacy policies, likely increasing the transparency (Degeling et al. 2019; Linden et al. 2019). These findings suggest that GDPR likely increased user privacy in terms of tracker intrusiveness and transparency. On the other hand, as we will elaborate in what follows, these requirements affect companies' operations, which may lead to economic loss. Moreover, companies' attempts to recoup these losses may have negative societal effects. For example, a company might scale back its services or charge for services once provided for free, resulting in a less-informed citizenry. Moreover, some companies might cut jobs, causing financial distress to their employees; if such layoffs take place on a large scale, the societal harm could be profound.

Thus, in establishing privacy regulations, policymakers must carefully balance between ensuring citizens' right to privacy and avoiding excessive damage to the performance of companies that use user data, given the potential societal effects of such damage. Yet, it is challenging to predict how implementing data privacy laws will affect companies' performance and revenue. Part of the challenge stems from users responding in unexpected ways to efforts to protect their privacy. Indeed, though users claim to value their privacy, it is well established that their actual behavior online does not necessarily align with these stated preferences (known as the privacy paradox; e.g., Acquisti 2004).

Accordingly, we present an empirical study to examine how the GDPR coming into effect, which we refer to as "enforcement of GDPR", affected user behavior on thousands of websites. We focus on two classes of user behavior metrics: user quantity (e.g., numbers of total visits) and usage intensity (e.g., page impressions per visit). These metrics are of interest as indicators of company performance. They often link with companies' revenues (e.g., e-commerce sites or sites with ad-based revenue; see the concluding sections of this paper).

Our analysis builds upon the premise that enforcing a privacy law can positively and negatively affect user quantity and usage intensity. Regarding user quantity, limitations on data collection and usage restrict companies' marketing activities, such as targeting new customers through personalized ads. As a result, users might be less aware of certain companies than they would have been otherwise and face increased search costs to find them. Consequently, traffic to those companies' websites might decrease. At the same time, traffic to certain websites might increase among users who find themselves with fewer alternatives—indeed, shortly after the enforcement date of GDPR, some websites operating outside the EU blocked access to EU-users to avoid having to comply with the law (Lecher 2018).

Regarding usage intensity, the requirements for transparency and consent to collect data may require websites to adjust their appearances—thereby affecting the user experience. For example, users might face a pop-up with information regarding the website's cookie usage or other data collection activities and then have to click to accept or decline cookies and the respective data collection. This interaction might increase users' awareness of their data disclosure and influence their usage intensity (Dinev and Hart 2006). In particular, they might spend less time on the website to reduce the amount of data it can collect, or they might abandon the website to avoid having to authorize it to collect data. Alternatively, once users have consented to have their data collected, they might use the website more than they would otherwise—to avoid having to visit other websites and authorize them to collect data. Lastly, there might be users who do not change their behavior at all.

These arguments suggest that, overall, the enforcement of a privacy law such as the GDPR, i.e., the GDPR becoming effective, may have positive or negative effects, or no effect at all, on the number of users who visit a particular website and on their usage intensity. Moreover, different websites might be affected differently, as users' expectations regarding their privacy and their consequent responses to privacy-driven changes in website operations may vary across regions (cultures) or websites in different industries (e.g., Dinev et al. 2006). It is also essential to understand how these effects develop over time, as it might take users several months to adjust their usage habits.

Thus, our study aims to achieve the following specific objectives:

1) Quantifying the effects of the enforcement of the GDPR on five metrics of user quantity and four metrics of usage intensity on websites over time (from 3 months up to 18 months after the enforcement of the GDPR);

2) Identifying how these effects vary as a function of website characteristics (i.e., website industry and popularity) and user characteristics (i.e., a user's country of origin).

Our analysis relies on a dataset capturing user behavior on 6,286 unique websites spanning 24 industries; these websites represent the most popular websites in 13 countries (11 EU countries, Switzerland, and the United States). The data cover the period from July 2017 to December 2019—i.e., 10 months before and 18 months after the enforcement of the GDPR (hereafter referred to as "GDPR") on May 25th, 2018—enabling us to construct a before-and-after analysis.

Within our dataset, some website–user interactions are subject to the GDPR (i.e., interactions involving EU-websites or EU-users). In contrast, others are not (i.e., interactions involving Non-EU-websites and Non-EU-users), effectively creating a "control group." Thus, we can use a panel differences estimator similar in spirit to a difference-in-differences (DiD) estimation (e.g., Janakiraman et al. 2018, Kumar et al. 2016, Goldstein et al. 2014). We combine the panel differences estimator with a synthetic control group approach (Abadie et al. 2015) to isolate the effect of the GDPR on our metrics of interest.

We obtain the following results:

1) Among websites to which the GDPR is applicable, the average number of visits per website decreases by almost 5% in the short-term and about 10% in the long-term; about two-thirds of websites continue to be negatively affected by the GDPR in the long-term. We similarly observe short-term decreases of 0.8%-3% in the average number of unique visitors, page impressions, and amount of time on the website, and long-term decreases of 6.6%-9.7%.

2) Among websites that suffer from a reduction in user quantity, the remaining users exhibit an increase in usage intensity—for example, the number of visits per user increases, on average, by about 4.8% at 18 months post-GDPR. Conversely, among websites that gain users after the GDPR, usage intensity decreases; e.g., the number of visits per user decreases, on average, by about 9.1% at 18 months post-GDPR.

3) The effects of the GDPR vary across websites; for example, less-popular websites lose more total visits (10%-21% drop) than more-popular websites (2%-9% drop), suggesting that the GDPR increases market concentration. The effects also vary across industries, with Entertainment and Leisure websites being most negatively affected (-12.5 to -13.8% after 18 months). In contrast, Business and Consumer Service websites even experience a positive effect (+4.7% after 18 months).

4) User characteristics (i.e., a user's country of origin) have only a small effect on how the GDPR affects user behavior.

## KNOWLEDGE ON EFFECTS OF PRIVACY CHANGES ON ONLINE USER BEHAVIOR

We draw from and contribute to two main streams of literature. Through surveys and lab experiments, the first stream attempts to illuminate users' attitudes towards data privacy and their responses to different levels of privacy or control over their data. The second stream uses field studies to examine the effects of privacy laws on various outcomes of interest.

*User Attitudes and Behavior with Regard to Privacy*

Lab experiments and survey-based studies have examined how users' attitudes and website usage behavior are affected by websites' handling of user privacy. The results of these studies point to a nuanced relationship between privacy and user behavior. For example, several studies based on consumer surveys suggest that when users perceive themselves as having more control over their privacy—specifically, more options to regulate their privacy—they experience lower privacy concerns (Martin 2015), a higher level of trust in a website, an increase in purchase intentions (Martin et al. 2017) and their willingness to disclose data to websites (Brandimarte et al. 2013; Acquisti et al. 2013; Malhotra et al. 2004; Culnan and Armstrong 1999). They can even react more positively to personalized ads (Tucker 2013).

Other studies, in contrast, find that different privacy levels do not affect user behavior: For example, Belanger and Crossler (2011) show that users share data with companies despite privacy concerns. This result may have been induced by users' feelings of powerlessness regarding their privacy (Few 2018). Acquisti et al. (2012) further show that user privacy concerns and preferences for the same level of privacy are not stable. The willingness to disclose data can depend on other factors like the amount and order of such data requests. These findings align with the privacy paradox, indicating that users' stated privacy preferences often differ from their actual behavior (e.g., Acquisti 2004).

Still, other studies suggest that including more privacy control options for users might negatively affect website usage. In particular, privacy features, such as requesting users' explicit consent for data collection and more transparency (as required by GDPR), can make users aware of data disclosure that they were not previously aware of (Dinev and Hart 2006), increase privacy concerns and thus reduce ad effectiveness (Kim et al. 2018). This awareness may lead users to feel warier about using the site and thus diminish their usage.

Dinev and Hart (2006) proposed the privacy calculus theory, which provides a framework encompassing all these different responses to privacy controls. Specifically, the theory suggests that the extent to which a user values privacy on a particular website depends on the user's privacy concerns, the user's trust in the website, and the value that the user derives from the website's offerings. Users with higher privacy concerns or lower trust towards a website may be more likely than others to respond favorably to more stringent privacy measures. In turn, when users attribute a high value to the website's offerings, they may be willing to sacrifice privacy in exchange for convenient access to those offerings and thus may be indifferent to privacy levels—or even respond unfavorably if privacy hurts the website's accessibility.

This theory suggests that users' responses to changes in a website's handling of privacy may vary across users and websites. Indeed, several studies show that differences in privacy perceptions and expectations depend on a user's country and cultural background (e.g., Dinev et al. 2006; Steenkamp and Geyskens 2006; Miltgen and Peyrat-Guillard 2014) and on the device used by a user to access a website (Melumad and Meyer 2020). The current study extends these findings by comparing how users in different countries vary in their responses to privacy laws and by considering variations across websites with different characteristics.

*Field Studies: Effects of Privacy Laws on Various Outcomes*

The findings outlined above suggest that it is likely to be challenging to predict how large populations of users will respond to the enforcement of new privacy laws. Accordingly, several studies use field data to construct event studies of users' revealed behavior after enforcing such laws. Examples that predate the GDPR are the work of Goldfarb and Tucker (2011a), who show that implementing the EU Privacy and Electronic Communications

Directive reduces ad effectiveness on websites, making it more challenging for ad-financed websites to generate revenues, and of Campbell et al. (2015) who show that privacy laws especially hurt smaller online companies. At the same time, Goldfarb and Tucker (2011b) further show that irrespective of privacy laws, ad effectiveness can diminish for strongly obtrusive and targeted ads, suggesting a positive effect of privacy laws on user welfare.

Several recent studies have specifically sought to characterize various effects of the GDPR. Some of these works focus on websites' actions in response to the law, showing that many update their privacy policies (Degeling et al. 2019) and increase their privacy policy length (Linden et al. 2019). Furthermore, an apparent reduction in third-party cookies occurs (Libert et al. 2018; Hu and Sastry 2019). Partly due to the anticipated reduction in third-party cookies, Mirreh (2018) predicts that websites could lose almost half of their traffic because of an inevitable shift of retargeting strategies, making it more challenging for companies to get users to their websites.

A study that is particularly relevant to our research is that of Goldberg et al. (2021), who measure how the GDPR affected recorded web traffic and e-commerce sales four months after the enforcement of the regulation. The authors show an average 11.70% drop in recorded page views from EU-users (Goldberg et al. 2021). Our empirical study delivers insights that greatly extend Goldberg et al.'s research. Primarily, our study adopts a long-term orientation for a substantially larger website sample, providing a more comprehensive analysis of GDPR. Given that the GDPR was the first major new privacy law in the EU since the e-Privacy Directive in 2002, users may have needed some time to adjust their behavior to the GDPR. Therefore, the full effect of the privacy law might only become observable after some time. Furthermore, our study examines differences in the effects across websites and users. Finally, the data sample of our study enables an empirical estimation of metrics covering actual traffic, whereas Goldberg et al.'s available data only allow an examination of recorded traffic. As the authors mention in their study, a change in recorded traffic after GDPR is, in fact, a combination of two changes: A change in the number of consenting users and a change in the actual traffic that these consenting users generate.

## DESCRIPTION OF EMPIRICAL STUDY

Our empirical study aims to analyze the effects of the enforcement of the GDPR on online user behavior, as reflected in measures of user quantity and usage intensity; to understand how these effects evolve over time (distinguishing between short-term effects – 3 months after enforcement –, up to long-term effects – 18 months after enforcement); and to reveal how these effects vary as a function of website and user characteristics.

### Background on the GDPR

The GDPR, which came into effect on May 25th, 2018, is the first major privacy law in Europe since the e-Privacy Directive in 2002. The GDPR regulates any activity performed on personal data from users located in the EU. As a regulation, the law is further binding for all websites based in EU countries; according to Article 3 of the GDPR, a website's "base" (and thus the applicability of the GDPR) is determined according to the geographical location where the website's data processing takes place. Websites within the scope of GDPR that do not comply with the privacy law face significant fines of up to 4% of the website's global annual turnover or €20 million, depending on the severity of the infringement.

The GDPR handles various privacy aspects that can affect how a user engages with a website. Similar to other approved or enforced privacy laws such as Brazil's Lei Geral de Protecao de Dados Pessoais (LGPD), India's Personal Data Protection Bill (PDPB), or Thailand's Personal Data Protection Act (PDPA), the GDPR has stringent privacy protection requirements (Lucente and Clark 2020). For example, the mentioned privacy laws all require websites to obtain a user's explicit consent for data processing like the GDPR, i.e., they all follow an opt-in approach for consent. Given the similar nature of GDPR compared with other privacy laws, the findings of this study likely mirror the effects of other privacy laws on user quantity and usage intensity on websites. At the same time, for privacy laws that are less strict than GDPR, such as the California Consumer Privacy Act (CCPA; Lucente and Clark 2020), the findings of this study might serve as an upper bound of the effects.

*Description of Set-Up of Empirical Study*

Before the GDPR becoming effective, users could not anticipate how websites would react to the diverse set of requirements imposed by GDPR. In our empirical study, we examine the effect of the enforcement of GDPR, i.e., the GDPR coming into effect, on the user behavior on websites. Most likely, websites complied with GDPR to different degrees. So, we do not measure the effect if all websites behaved entirely according to GDPR. Instead, we observe the effect of the websites' interpretation of the privacy law. Thus, we measure what happened after GDPR came into effect – the intention-to-treat effect of GDPR. Therefore, our treatment "enforcement of GDPR" refers to "GDPR coming into effect" (on May 25, 2018) and not to a situation in which GDPR was enforced such that all websites behaved entirely with GDPR.

The GDPR provides a useful setting for quantifying the effect that the enforcement of privacy laws has on user behavior because it implicitly divides website–user interactions (here referred to as "observations") into a treatment group (i.e., GDPR is applicable) and a control group (i.e., GDPR does not apply), as depicted in Figure 1. As noted above, the GDPR's scope includes all websites based in the EU and further encompasses the processing of personal data from all users located in the EU. Thus, the treatment group comprises observations corresponding to EU-users visiting any website or to Non-EU-users visiting EU-websites. The control group consists of observations corresponding to Non-EU-users visiting Non-EU-websites. In line with Article 3 of the GDPR, we use the website's server location (retrieved from https://check-host.net) to determine the respective website's data processing location and the GDPR's applicability. We use the enforcement date of GDPR (May 25[th], 2018) to construct a before-and-after analysis, comparing the treatment group to the control group to quantify the intention-to-treat effect of GDPR. This approach allows us to construct a panel differences estimator that is similar in spirit to a DiD estimator and rests upon two critical assumptions: the stable unit treatment value assumption (SUTVA) and the parallel pre-treatment trends of the control and the treatment group.

*Figure 1: Scope of GDPR and Resulting Assignment to Treatment and Control Group*

| | | Base of User | |
|---|---|---|---|
| | | EU | Non-EU |
| Base of Website | EU | GDPR applies = Treatment Group | GDPR applies = Treatment Group |
| | Non-EU | GDPR applies = Treatment Group | GDPR does not apply = Control Group |

☐ Treatment Group    ☐ Control Group

Several factors might bias the treatment effect that we observe using the described methodology. For example, there might be concerns regarding the possible late or early compliance of websites with GDPR or the potential existence of confounding factors. The major concern, however, might be regarding the validity of our control group. This concern stems from the possible situation that websites in our control group might voluntarily comply with GDPR. Furthermore, the mere knowledge of Non-EU-users about the GDPR already represents a "treatment" that affects our control group as well. Both situations would represent a violation of the SUTVA that is integral to our analysis.

We thoroughly examine the robustness of our results to all of those factors, i.e., late or early compliance, confounding factors, and the possibility that GDPR also treats our control group. All robustness checks indicate that the mentioned factors do not bias our results (see Web Appendices D, E, G). Thus, even if a potential bias existed within our results, its impact is likely relatively small. Furthermore, such an effect only yields to underestimating GDPR's actual effect because the treatment might also impact the control group.

*Overview of Data*

*Description of data sample.* This study utilizes data from SimilarWeb for the Top 1,000 websites—as listed in Alexa Top Sites in April 2018—of two Non-EU countries (Switzerland and USA) and 11 EU countries (Austria, Denmark, France, Germany, Hungary, Italy, Netherlands, Poland, Spain, Sweden, and the UK[1]). The authors choose the USA and Switzerland as both countries are culturally similar to the EU. SimilarWeb draws on a diversified and rich global user panel to measure online user behavior. Companies (e.g., Google, Alibaba, eBay, P&G) primarily use data from SimilarWeb, but also researchers in top-tier academic journals (e.g., Calzada and Gill 2020, Lu et al. 2020). The websites in our sample span diverse industries (see Figure W 1 in Web Appendix A), audiences, and popularity levels (here measured by SimilarWeb ranks). For each website in our sample, the dataset also includes information about the website industry as well as the global, country, and industry rank, based on the website's popularity worldwide, in the analyzed country, and within the website's industry.

---

[1] During the time of our study, the UK was still a member of the EU. Its membership ended on January 31[th], 2020.

For each website in the sample, the dataset includes information on the user quantity metrics of users accessing that website from one EU country: the one in which the website is most popular. Additionally, for each website, user quantity data are available for users accessing the website from the US. Thus, if a website does not appear in the Top 1,000 of any EU country, data are available only for US users. These data span the period between July 1$^{st}$, 2017 and December 31$^{st}$, 2019—i.e., almost a year before GDPR's enforcement (May 25$^{th}$, 2018) and 1.5 years after the enforcement—and can therefore be used for a before-and-after analysis as outlined above.

We start with 13 countries with 1,000 websites each. Our initial sample includes 7,332 unique websites after we removed duplicate websites. For example, "google.com" is a duplicate website as it is among the Top 1,000 websites in all 13 countries. Instead of occurring 13 times, google.com just occurs once in our sample. For each of these 7,332 websites, we have user behavior data corresponding to Non-EU-users. For 6,460 websites of those 7,332 websites, the dataset additionally includes user behavior data of EU-users.

Thus, for 6,460 websites, we have two sets of observations, corresponding, respectively, to the Non-EU-user base and to the EU-user base of that website. For the remaining 872 websites, we only observe the Non-EU-user base. In what follows, we consider each website's Non-EU and EU-user bases separately and refer to each combination of a website with one of the two user bases, for convenience, as a "website-instance." For example, for a website such as "zeit.de" that is based in an EU country (here, Germany), we observe two website-instances: One website-instance corresponds to the set of observations for the EU-user base of "zeit.de." The second website-instance corresponds to the set of observations for the Non-EU-user base of "zeit.de." As "zeit.de" is EU-based, GDPR applies to both its website-instances, and both website-instances belong to the treatment group (Figure 1).

Accordingly, for a website such as "nzz.ch" that is based in a Non-EU country (here: Switzerland), we observe two website-instances: one website-instance corresponding to the set of observations for the Non-EU-user base of "nzz.ch" and the second website-instance corresponding to the set of observations for the EU-user base. As the website of this second example is not EU-based, GDPR applies only to the website-instance that corresponds to the EU-user base of "nzz.ch," which belongs to the treatment group, but not to the Non-EU-user base, which belongs to the control group (see Figure 1).

Overall, the initial sample includes 7,332 website-instances corresponding to a set of observations of the Non-EU-user base and 6,460 website-instances corresponding to a set of observations of the EU-user base, totaling 13,792 website-instances. We then drop website-instances for which the user base generated, on average, fewer than 1,000 visits per week or not a single visit for more than an entire month in the observation period. We also drop website-instances that exhibited strong traffic drops or peaks at some point in time that our available data cannot explain. Especially the website-instances that include visits to EU-websites from Non-EU-users exhibit a low average number of visits due to many EU-websites not being popular in Non-EU countries. This procedure results in a final sample of 9,683 website-instances, corresponding to 6,286 unique websites (Table 1).

For 3,397 websites, we have two website-instances (EU and Non-EU-user bases), and for 2,889 websites, we have one website-instance (EU or Non-EU-user base). Overall, as we also show in Table 2, 5,683 websites, corresponding to 7,982 website-instances, belong to our treatment group, encompassing over 1.15 trillion total website visits from the EU.

*Table 1: Derivation of Final Sample of Website-Instances*

| | Website-Instances with EU-user data | Website-Instances with Non-EU-user data |
|---|---|---|
| **Sample** of (non-unique) websites (top 1,000 websites of 11 EU countries, CH and US) | 11,000 | 12,000 |
| **Sample after removal** of duplicated and non-existent websites (e.g., fraudulent pop-ups) | 6,460 | 7,332 |
| **Sample after additional removal** of website-instances with average weekly visits <1,000 | 6,308 | 5,407 |
| **Final sample** after additional removal of website-instances with no visits in >1 month or strong outliers | **5,494** (3,643 EU websites, 1,851 Non-EU websites) | **4,189** (2,488 EU websites, 1,701 Non-EU websites) |
| | Total number of websites: **6,286** | |
| **Final sample (only for unique visitor analysis)** after additional removal of website-instances with monthly unique visitors <5,000 in at least one month | **5,105** | **3,103** |
| *Note: SimilarWeb does not report unique visitor information for websites with < 5,000 unique visitors. Thus, the final sample for the unique visitor analysis contains only website-instances that had > 5,000 unique visitors throughout the entire observation period.* | | |

SimilarWeb does not report unique visitor information for websites with less than 5,000 unique visitors in a month. Therefore, the unique visitor analysis contains a smaller set of 5,198 treated websites. Our control group consists of 1,701 websites, corresponding to the same number of website-instances (see Table 2), encompassing almost 1.8 trillion total website visits from Switzerland and the US.

*Table 2: Distribution of Website-Instances in Treatment and Control Groups*

| | | Base of User | | |
|---|---|---|---|---|
| | | EU | Non-EU | |
| Base of Website | EU | 3,643 website-instances | 2,488 website-instances | ≙ 6,131 website-instances, corresponding to 3,832 EU-based websites |
| | Non-EU | 1,851 website-instances | 1,701 website-instances | ≙ 3,552 website-instances, corresponding to 2,454 Non-EU-based websites |
| | | ≙ 5,494 website-instances, corresponding to 5,494 websites | ≙ 4,189 website-instances, corresponding to 4,189 websites | |

☐ Treatment Group    ☐ Control Group

*Description of user quantity and usage intensity metrics.* In what follows, we define our variables of interest, namely, our user quantity and usage intensity metrics. The examined variables are connected to some extent. Specifically, all metrics correspond with the weekly total number of visits, which is our main metric of interest. Still, despite the connectivity between the user quantity metrics, each provides a slightly different insight into the effects of the GDPR on user behavior. Furthermore, examining the user quantity metrics in relation to our main metric enables an examination of the different usage intensity metrics on websites.

We, therefore, compare the respective effects of GDPR on the user quantity metrics for each website to examine GDPR's effect on the usage intensity metrics. More specifically, we calculate the effect of GDPR on the usage intensity metrics based on the changes of the respective user quantity metrics corresponding to a specific usage intensity metric. For an overview of the relationship between the metrics, see Table 3.

*Table 3: Relationship between the User Quantity and Usage Intensity Metrics*

| User Quantity Metric | Corresponding Formula | Usage Intensity Metric |
|---|---|---|
| **Weekly Total Number of Visits** | Main Metric | |
| *Number of visits to a website in a week* Measure of website's traffic volume | Corresponds to all usage intensity metrics | |
| **Monthly Number of Unique Visitors** | Visits per Unique Visitor | **Visits per Unique Visitor** |
| *Number of unique users visiting a website in a month* Measure of website's reach | $= \dfrac{\text{Total Number of Visits}}{\text{Number of Unique Visitors}}$ | *Average number of visits per unique visitor* |
| **Weekly Number of Page Impressions** | Page Impressions per Visit | **Page Impressions per Visit** |
| *Number of pages visited per week on a website by the entire user base* Measure of website's ability to spark engagement | $= \dfrac{\text{Number of Page Impressions}}{\text{Total Number of Visits}}$ | *Average number of pages viewed per visit* |
| **Weekly Time on Website** | Time per Visit | **Time per Visit** |
| *Time in minutes spent in a week on a website by the entire user base* Measure of website's ability to spark interest | $= \dfrac{\text{Time on Website}}{\text{Total Number of Visits}}$ | *Average time spent on a website per visit* |
| **Weekly Number of Bouncing Visitors** | Bounce Rate | **Bounce Rate** |
| *Number of visits to a website in a week in which the user views only one page* Measure of website's ability to retain traffic | $= \dfrac{\text{Number of Bouncing Visitors}}{\text{Total Number of Visits}}$ | *Share of visitors leaving a website after just one page* |

We analyze all user quantity metrics weekly, except for the number of unique visitors, for which data are only available monthly. Due to large differences in the values of each metric across websites and countries, we convert all user quantity metrics (+1 to avoid zero values) to their natural logarithm so that we capture relative (i.e., percentage) effects. Figure 2 depicts the mean comparison (before and after GDPR) of the log-transformed user quantity variables for the 7,892 treated website-instances and the 1,701 control website-instances.

We calculate the effects for the user quantity metrics per website-instance. We then determine the effect of GDPR on a website as follows: If only one website-instance corresponds to a specific website, i.e., only data for one user base is available for that website, the effect of GDPR on that website comprises only the effect of that one website-instance. All Non-EU-websites (for these websites, only one website-instance belongs to the treatment group, see Figure 1) and about 30% of the EU-websites correspond with only one website-instance.

*Figure 2: Comparison of Logarithm of User Quantity Metrics (i.e., Dependent Variables) between Treatment and Control Group and Pre- and Post-GDPR Period*

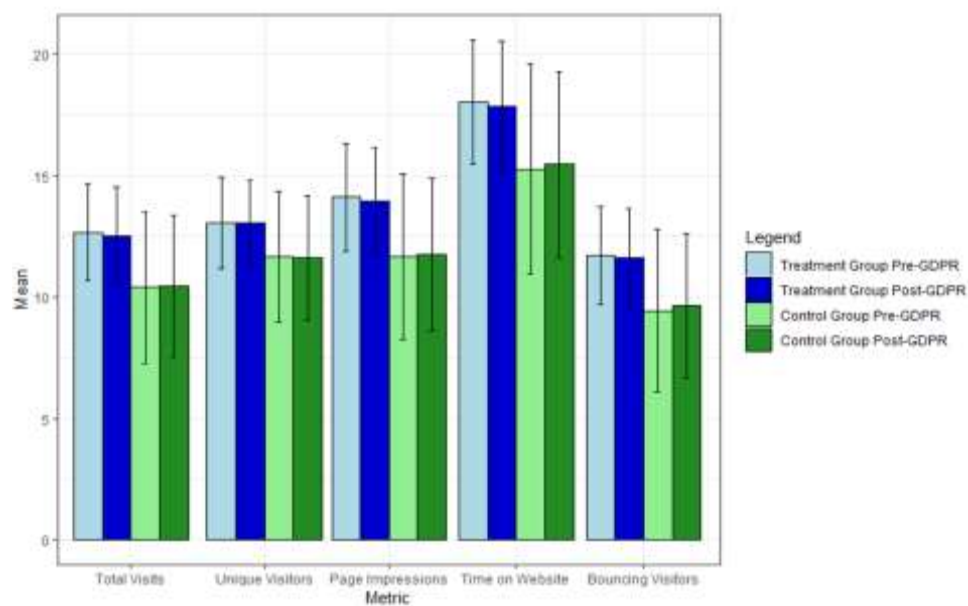

For 70% of the EU-websites, two treated website-instances correspond to the same website, i.e., data for the EU and Non-EU-user base are available. For these websites, the overall effect of GDPR comprises the effects of both website-instances. Hereby, we consider the relative sizes of the two website-instances before GDPR when merging the two effects into one. For example, the website "zeit.de," a reputable German online news website, received 98.94% of its visits from German (EU) users. Thus, GDPR's effect on the number of visits on the website "zeit.de" comprises 98.94% of its effect on the website-instance corresponding to the EU-user base of "zeit.de," and 1.06% of GDPR's effect on the website-instance corresponding to the Non-EU-user base of "zeit.de." This weighting procedure results in the same effects as we would have determined if we had combined the two website-instances from the beginning of the calculation.

We then compare the respective effects of GDPR on the user quantity metrics for each website to examine GDPR's effect on the usage intensity metrics that are calculated based on the changes of the user quantity metrics.

*Description of Methodology to Analyze Data*

We do not observe a control group for some website-instances, primarily due to the GDPR affecting both website-instances for a large share of our websites. Thus, we examine the effect of the GDPR on the various user behavior metrics using a combination of a panel differences estimation and the synthetic control group (SCG) approach and exploit the enforcement of the GDPR as the "treatment" event. The panel differences estimation is similar in spirit to a DiD approach (e.g., Janakiraman et al. 2018, Kumar et al. 2016, Goldstein et al. 2014) and aims to isolate the effect of treatment by comparing the differences before and after treatment (here GDPR) across two panels (here the treatment and control group). Combining the two approaches enables us to examine the differential impact that the enforcement of GDPR had on user behavior compared to our synthetic control group.

*Description of methodology to analyze user quantity metrics.* Using the regression formula below and the control and treatment group assignment described in Figure 1, we calculate the treatment effect ($\beta_{3,q,wi}$) for every website-instance *wi* for each user quantity metric *q*. To determine the development of the treatment effect over time *t* (here measured in weeks for all user quantity metrics except unique visitors where we measure *t* in months), we rerun our analysis several times, extending the duration of the post-GDPR observation period in each analysis. We first consider a post-treatment period of 3 months after GDPR (up to August 25th, 2018, thus including observations from week 1 to week 60), then periods of 6 (week 1 to 73), 9 (week 1 to 86), 12 (week 1 to 99) and 18 (week 1 to 125) months.

These analyses enable us to determine the GDPR's short- up to the long-term effects.

(1) $ln\big(Y_{q,t,wi} + 1\big) = \beta_{0,q,wi} + \beta_{1,q,wi} * EU_{wi} + \beta_{2,q,wi} * Postperiod_t + \beta_{3,q,wi} * Treated_{t,wi} + \epsilon_{q,t,wi}$

| | |
|---|---|
| $Y_{q,t,wi}$: | Value of user quantity metric *q* in week *t* on website-instance *wi* |
| $EU_{wi}$: | EU-Dummy, i.e., binary variable for which a value of 1 indicates that the users or website of website-instance *wi* are EU-based, else 0 |
| $Postperiod_t$: | Postperiod-Dummy, i.e., binary variable for which a value of 1 indicates that the observation in week *t* lies in the post-treatment period, else 0 |
| $Treated_{t,wi}$: | = $EU_{wi} * Postperiod_t$; Treatment-Dummy, i.e., binary variable for which a value of 1 indicates that in week *t*, website-instance *wi* needs to consider GDPR, otherwise 0 |
| $\epsilon_{q,t,wi}$: | Error term for user quantity metric *q* in week *t* for website-instance *wi* |

We rely on the SCG method (Abadie et al. 2015), which entails a synthetic construction of a control group whose pre-treatment patterns are comparable to those of the treatment group. We construct this matched control group by selecting, for each treated website-instance, a weighted combination of several control website-instances. Thus, this approach requires (i) choosing a set of control website-instances to use and (ii) weighing each website-instance. We weigh such that the weighted combination of control website-instances, referred to collectively as the "synthetic control website-instance," minimizes the pre-treatment mean squared error (MSE) between the resulting synthetic control website-instance and the treated website-instance (following the approach outlined in Xu (2017)). Thus, this approach fulfills the parallel pre-treatment condition by construction. Then, we calculate the post-treatment metric of interest for the synthetic control website-instance that serves as the treated website-instance's counterfactual.

For each metric, we follow Abadie (2021) and Abadie et al. (2015) and carefully choose the control website-instances to obtain a reasonable control for the treated website-instance. The set of controls should (i) avoid the risk of overfitting, i.e., not be too large, and (ii) avoid the risk of bias, i.e., not exhibit large differences in (un-)observed factors compared to the treated website-instance. Thus, we select (i) five website-instances that (ii) belong to the same industry as the treated website-instance and (iii) have the highest pre-treatment correlations with the respective metric of the treated website-instance.

Using these five control website-instances, we follow the approach outlined above to calculate the weights of these website-instances, to create a synthetic control website-instance that exhibits a similar pre-treatment pattern as the treated website-instance. We then use the weights and observed values of the five website-instances to calculate a synthetic time series for the synthetic control website-instance, spanning the post-treatment period. The outcomes of these calculations serve as the control group to determine the effect of the treatment on the metric of interest for all website-instances. We repeat the process for each user quantity metric $q$ and website-instance $wi$.

We then determine the effect of GDPR on a website, referred to as $\Delta$, as described above: If only one website-instance corresponds to a website, the GDPR's effect on that website-instance and user quantity metric ($\beta_3$ calculated in Equation 1) determines the GDPR's effect on that website (Δ). If two website-instances correspond to a website, the effects of GDPR on both website-instances for the user quantity metric (two treatment effects $\beta_{3,wi}$) determine the GDPR's effect on that website (Δ), taking the relative sizes of the two website-instances for that one website into account. We determine these relative sizes of the two website-instances for that one website as an average of the entire pre-treatment period.

*Description of methodology to analyze usage intensity metrics.* After these steps, we have determined the effect of interest, the treatment effect $\beta_3$, for all treated websites-instances, user quantity metrics, and post-treatment periods, and merged the treatment effects of the website-instances to obtain the treatment effects Δ for all corresponding websites and user quantity metrics. We then use these treatment effects for all websites and post-treatment periods to examine the change in our usage intensity metrics for each website over time. For this examination, we take advantage of two aspects: First, each usage intensity metric is a function of two user quantity metrics, as shown in Table 3. For example, the number of visits per unique visitor is a function of the number of unique visitors and the total number of visits on a website *w* (see Table 3):

$$(2)\ Number\ of\ Visits\ per\ Unique\ Visitor_w = \frac{Total\ Number\ of\ Visits_w}{Number\ of\ Unique\ Visitors_w}$$

Second, the treatment effects calculated with Equation (1) are relative (i.e., approximately percentage) changes of our user quantity metrics for each post-treatment period *p*. Thus, to visualize the relative change of the number of visits per unique visitor due to GDPR for a particular website *w* for a particular post-treatment period *p*, we include the GDPR's effect in Equation (2):

$$(3)\ Number\ of\ Visits\ per\ Unique\ Visitor_w * \left(1 + \Delta Number\ of\ Visits\ per\ Unique\ Visitor_{p,w}\right) =$$

$$\frac{Total\ Number\ of\ Visits_w * (1 + \Delta Total\ Number\ of\ Visits_{p,w})}{Number\ of\ Unique\ Visitors_w * (1 + \Delta Number\ of\ Unique\ Visitors_{p,w})}$$

We use Equation (1) to calculate the GDPR's effect, reflected in $\Delta$, for the two user quantity metrics (number of unique visitors and total number of visits). To determine the effect on the usage intensity metric (number of visits per unique visitor), we rearrange Equation (3):

$$(4)\ \Delta Number\ of\ Visits\ per\ Unique\ Visitor_{p,w} = \frac{1 + \Delta Total\ Number\ of\ Visits_{p,w}}{1 + \Delta Number\ of\ Unique\ Visitors_{p,w}} - 1$$

| | |
|---|---|
| $\Delta Number\ of\ Visits\ per\ Unique\ Visitor_{p,w}$: | GDPR's effect in period $p$ on the number of visits per unique visitor for website $w$ |
| $\Delta Total\ Number\ of\ Visits_{p,w}$: | GDPR's effect in period $p$ on the total number of visits for website $w$ |
| $\Delta Number\ of\ Unique\ Visitors_{p,w}$: | GDPR's effect in period $p$ on the number of unique visitors for website $w$ |

This process (see Web Appendix B for a detailed derivation of Equations 3 and 4) enables us to reveal the GDPR's effects on the number of visits per unique visitor for each website (within the final sample for the unique visitor analysis) and each post-treatment period $p$ (i.e., after 3, 6, 9, 12, and 18 months of GDPR). We calculate the GDPR's effects on the other usage intensity metrics $i$ (i.e., page impressions per visit, time per visit, and bounce rate) with the same procedure. However, these usage intensity metrics are a function of the total number of visits and different user quantity metrics $q$ (see Table 3). Therefore, we need to adjust Equation (4) slightly:

$$(5)\ \Delta Usage\ Intensity\ Metric_{i,p,w} = \frac{1 + \Delta User\ Quantity\ Metric_{p,q,w}}{1 + \Delta Total\ Number\ of\ Visits_{p,w}} - 1$$

| | |
|---|---|
| $\Delta Usage\ Intensity\ Metric_{i,p,w}$: | GDPR's effect on the usage intensity metric $i$ (Page Impressions per Visit, Time per Visit, Bounce Rate) in period $p$ for website $w$ |
| $\Delta User\ Quantity\ Metric_{p,q,w}$: | GDPR's effect in period $p$ on the corresponding user quantity metric $q$ (Number of Page Impressions, Time on Website, Number of Bouncing Visitors) for website $w$ |
| $\Delta Total\ Number\ of\ Visits_{p,w}$: | GDPR's effect in period $p$ on the total number of visits for website $w$ |

*Description of methodology to analyze variations of effects across websites.* After calculating the GDPR's effects on usage quantity and usage intensity for each website, we subsequently classify the websites according to a particular feature of interest—namely, website industry, popularity (measured by the ranks within SimilarWeb's global, country, and industry rankings of websites), and the country of origin of the predominant user base—and examine whether specific website or user characteristics are associated with positive or negative as well as stronger or weaker effects due to GDPR.

## RESULTS OF EMPIRICAL STUDY

The following subsections outline the distribution of the GDPR's effect across websites for user quantity and usage intensity. GDPR does not affect all websites the same way. While some websites experience negative effects, others are not affected by GDPR or even experience positive effects. The size of GDPR's effects further differs across websites. As we later show, the GDPR's effects on the analyzed metrics result in significant economic effects for websites.

### GDPR's Effect on User Quantity Metrics

We visualize in Figure 3 the distribution of the GDPR's effect on our main metric, the weekly total number of visits, over time. Figure 3 includes all websites, irrespectively of whether GDPR affects them significantly.

On average, in the 3 months after the GDPR, treated websites experience an average decline in visits of about 4.88%. Over time, this decrease becomes even stronger: After 1.5 years of GDPR, the number of visits to the treated websites is 10.02% lower due to GDPR. We further find that 3 months after the GDPR, GDPR affects 59.31% of all websites negatively (the solid line plot in Figure 3 indicates the share of negatively affected websites). The share of websites that experience a decrease in visits increases to 66.70% after 18 months. For the rising share of websites that GDPR affects negatively, the corresponding negative effect becomes even more negative over time. At the same time, while there are websites that benefit from GDPR, particularly shortly after GDPR, the effect sizes corresponding to these positive effects decrease over time. For some of the initially positively affected websites, the effects even become negative after 12 or 18 months.

*Figure 3: Distribution of the Effect of GDPR on Weekly Total Number of Visits over Time*

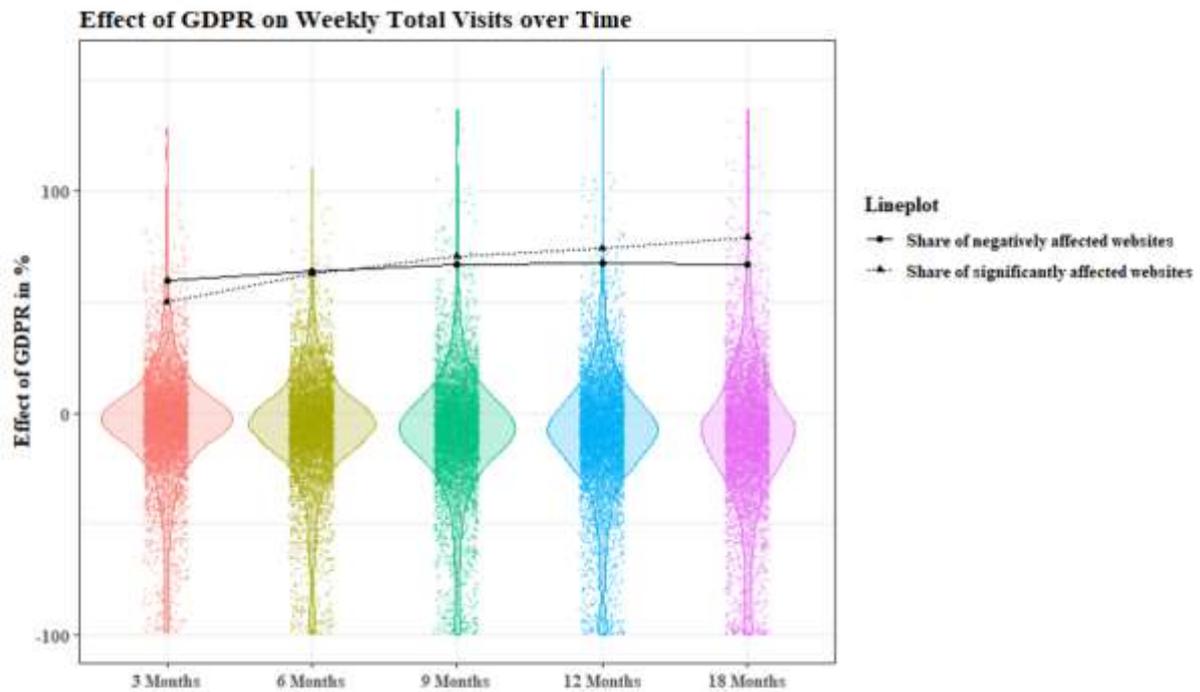

Still, not all websites experience significant (i.e., statistically different from zero) effects by GDPR. Three months after GDPR, only about half of the websites (49.93%) experience significant effects (the dashed line plot in Figure 3 indicates the share of significantly (on the 5%-level) affected websites). Still, the share of significantly affected websites rises to 78.83% after 18 months of the GDPR. Overall, after 18 months for total visits, GDPR has had a significant positive effect for 23.96%, a significant negative effect for 54.87%, and no significant effect for 21.17% of all websites.

Table 4 summarizes the GDPR's effect directions, the share of negatively affected websites, the share of significantly affected websites, and the development of the effects over time for our main user quantity metric, the total number of visits, and the four other user quantity metrics (see Web Appendix A for more details on the other metrics).

*Table 4: Summary of Results for User Quantity Metrics*

| Metric | | 3 months | 6 months | 9 months | 12 months | 18 months |
|---|---|---|---|---|---|---|
| **Total Visits** | Median: | -3.49% | -5.54% | -7.54% | -8.24% | -8.91% |
| | Mean: | -4.88% | -7.22% | -9.07% | -9.57% | -10.02% |
| | Share of significant effects: | 49.93% | 62.79% | 70.64% | 74.05% | 78.83% |
| | Share of negative effects: | 59.31% | 63.79% | 66.94% | 67.41% | 66.70% |
| **Unique Visitors** | Median: | -1.24% | -3.50% | -5.60% | -6.04% | -6.65% |
| | Mean: | -0.77% | -3.27% | -5.50% | -6.18% | -6.61% |
| | Share of significant effects: | 54.15% | 66.14% | 73.43% | 75.82% | 80.52% |
| | Share of negative effects: | 53.40% | 58.63% | 61.64% | 62.41% | 61.73% |
| **Page Impressions** | Median: | -2.75% | -3.92% | -5.44% | -6.04% | -9.29% |
| | Mean: | -3.12% | -4.83% | -6.33% | -6.48% | -9.28% |
| | Share of significant effects: | 46.53% | 58.56% | 66.21% | 70.51% | 76.06% |
| | Share of negative effects: | 56.25% | 58.69% | 60.06% | 60.61% | 64.19% |
| **Time on Website** | Median: | -4.51% | -6.07% | -8.42% | -9.19% | -9.50% |
| | Mean: | -4.72% | -6.84% | -8.89% | -9.87% | -9.68% |
| | Share of significant effects: | 43.88% | 55.71% | 64.30% | 67.72% | 74.18% |
| | Share of negative effects: | 59.46% | 61.84% | 64.89% | 65.95% | 63.55% |
| **Bouncing Visitors** | Median: | -4.14% | -6.77% | -8.94% | -9.48% | -10.16% |
| | Mean: | -4.35% | -7.28% | -9.48% | -9.94 % | -10.16% |
| | Share of significant effects: | 47.22% | 59.92% | 68.73% | 73.10% | 79.00% |
| | Share of negative effects: | 59.27% | 64.57% | 67.80% | 67.43% | 66.27% |

The table shows a summary of GDPR's effect on user quantity metrics. The table shows the mean and median values of the change in the metrics due to GDPR over all websites in each of the analyzed periods. The share of significantly different effects (on the 5%-level) from zero is reported for each period.
For example, the 3-month effect of GDPR for total visits over all websites (second row / third column) was on average -4.88%, the median effect was -3.49%, and 49.93% of the websites were significantly affected.

The results for our other four user quantity metrics are all substantively similar to our main user quantity metric, the total number of visits. For example, the GDPR affects the average treated website negatively for all user quantity metrics in all examined time points. Furthermore, the share of negatively affected websites increases over time (between 53.40% and 59.46% after 3 months; between 61.73% and 66.70% after 18 months), and the share of significantly affected websites increases over time (between 43.88% and 54.15% after 3 months; between 74.18% and 80.52% after 18 months).

The most prominent – albeit small – differentiating factor across the user quantity metrics is the size of the effects. Most notably, GDPR affects treated websites' page impressions, time on the website, and bouncing visitors similarly as total visits (on average -3.12% to -4.88% after 3 months; -9.28% to -10.16% after 18 months). Contrastingly, GDPR's effect on unique visitors is smaller than our main metric: Three months after GDPR, the number of unique visitors decreases by 0.77% and 6.61% after 18 months of GDPR.

*GDPR's Effect on Usage Intensity Metrics*

The differences in the effect sizes across the user quantity metrics reveal that while there are dependencies among the user quantity metrics, additional metrics influence the observed effect sizes – the usage intensity metrics (see Table 3 for the relationship between the user quantity and usage intensity metrics). If the usage intensity for an average user on a website had stayed the same post-GDPR, there would be no difference in the effect sizes for the different user quantity metrics. Thus, the differences in the user quantity metrics uncover that the underlying usage intensity changes due to GDPR, which we examine in what follows. Table 5 shows a summary of GDPR's effects on the four usage intensity metrics.

Unlike the user quantity metrics, which all exhibited the same average direction of effects, we observe different effect directions across the usage intensity metrics. For example, after 3 months of GDPR, there is a decrease in the average number of visits that a unique visitor conducts to a website (-1.62%), the average time per visit (-0.83%), and the bounce rate (+0.81%; note that an increase in the bounce rate is an undesired development). At the same time, the page impressions per visit increase post-GDPR (+2.44% after 3 months).

*Table 5: Summary of GDPR's Effect on Usage Intensity Metrics*

| Metric | | 3 months | | | 18 months | | |
|---|---|---|---|---|---|---|---|
| | | Share of Websites | Median Effect | Mean Effect | Share of Websites | Median Effect | Mean Effect |
| **Visits per Unique Visitor** | **All Treated Websites** | *100.00%* | -2.62% | -1.62% | *100.00%* | -2.81% | -0.59% |
| | **Treated Websites that Gain Unique Visitors** | *46.60%* | -6.19% | -6.46% | *38.27%* | -9.37% | -9.09% |
| | **Treated Websites that Lose Unique Visitors** | *53.40%* | +1.23% | +2.56% | *61.73%* | +1.39% | +4.77% |
| **Page Impressions per Visit** | **All Treated Websites** | *100.00%* | +0.56% | +1.97% | *100.00%* | +0.28% | +2.15% |
| | **Treated Websites that Gain Total Visits** | *40.69%* | -2.92% | -2.44% | *33.30%* | -4.87% | -4.58% |
| | **Treated Websites that Lose Total Visits** | *59.31%* | +3.07% | +5.05% | *66.70%* | +3.11% | +5.53% |
| **Time per Visit** | **All Treated Websites** | *100.00%* | -1.96% | -0.83% | *100.00%* | -0.93% | +0.09% |
| | **Treated Websites that Gain Total Visits** | *40.69%* | -3.92% | -3.18% | *33.30%* | -4.40% | -4.02% |
| | **Treated Websites that Lose Total Visits** | *59.31%* | -0.09% | +0.82% | *66.70%* | +1.25% | +2.14% |
| **Bounce Rate** | **All Treated Websites** | *100.00%* | -0.57% | +0.81% | *100.00%* | -0.58% | +2.51% |
| | **Treated Websites that Gain Total Visits** | *40.69%* | -3.36% | -2.86% | *33.30%* | -4.81% | -3.76% |
| | **Treated Websites that Lose Total Visits** | *59.31%* | +1.36% | +3.38% | *66.70%* | +1.36% | +5.67% |

The table shows a summary of GDPR's effect on the usage intensity metrics. The table shows the mean and median values of the change in the metrics due to GDPR 1) over all websites, 2) over the websites that experience positive user quantity effects, and 3) over the websites that experience negative user quantity effects in each of the analyzed periods. For example, the average 3-month effect of GDPR for visits per unique visitor over all websites (third row / fifth column) was -1.62%, and the median effect (third row / fourth column) was -2.62%.

We further observe that over time, the GDPR's effect becomes more positive: The initial average reduction in the time per visit becomes positive after 18 months of GDPR (+0.09%), the page impressions per visit increase even more after 18 months of GDPR (+2.15%), and the reduction of visits per unique visitor becomes weaker (-0.59%). However, the bounce rate increases over time (+2.51% after 18 months of GDPR).

Given the substantial differences in effect sizes and directions across websites for our user quantity metrics, we further examine the usage intensity by dividing the websites into groups that experienced an increase or decrease in the user quantity due to GDPR. Particularly, 46.60% and 38.27% of websites increased unique visitors after respectively 3 and 18 months; total visits increased for respectively 40.69% and 33.30% (Table 5).

Using this division, we observe that the effect direction of all usage intensity metrics aligns for the two groups: Websites that experience an increase in user quantity exhibit a decrease in usage intensity of the average user, and websites that experience a decrease in user quantity exhibit an increase in usage intensity. Thus, among websites that lose unique visitors, the remaining visitors visit those websites more often. However, websites that gain unique visitors gain visitors who re-visit less frequently and have a lower usage intensity for these visits in terms of page impressions and time.

The corresponding effects become even stronger over time for all usage intensity metrics, i.e., the positive effects become more positive, and the negative effects become more negative. For example, 3 months after GDPR, the websites that gain unique visitors experience a 6.46% decrease in the number of visits per unique visitor compared with pre-GDPR levels; 18 months after GDPR, unique visitors decreased by 9.09%. Yet, the websites that lose unique visitors experience a 2.56% increase in the number of visits per unique visitor after 3 months, and 4.77% after 18 months. We observe the same effect directions and similar – albeit smaller – effect sizes for the other usage intensity metrics: For websites that increase total visits, after 3 (18) months of GDPR, the average number of page impressions decreases by 2.44% (4.58%), the time per visit by 3.18% (4.02%), and the bounce rate by 2.86% (3.76%). For the websites that lose total visits, after 3 (18) months of GDPR, the average number of page impressions increases by 5.05% (5.53%), the time per visit by 0.82% (2.14%), and the bounce rate by 3.38% (5.67%).

*Variation in GDPR's Effects as a Function of Characteristics of the Website and the User*

The previous section examined the distribution of GDPR's effect across websites on the user quantity and usage intensity, showing that GDPR has affected websites differently. In what follows, we analyze how the effects of the GDPR on user quantity metrics vary as a function of website characteristics—website industry and website popularity—and user characteristics, namely, users' country of origin. For each of these analyses, we classify the websites according to the focal feature of interest (e.g., website industry) and calculate the average effect of GDPR on our main user quantity metric, i.e., the total number of visits, across all websites within each category (e.g., same industry).

*Variation in the GDPR's effects as a function of the industry of the website.* Figure 4 shows that GDPR affects websites from different industries in very different ways. Websites within the "Heavy Industry and Engineering" and "Gambling" industries show the most negative effects, losing an average of almost 50% and 20% of their total visits 3 months after GDPR, followed by "Lifestyle," "Games," "Arts and Entertainment," "Reference Materials" and "Hobbies and Leisure." Websites in the "Business and Consumer Services" and "Vehicles" industries experience positive effects throughout the entire observation period. Some industries exhibit positive effects shortly after the GDPR and subsequently experience negative effects, such as "Travel and Tourism" and "E-Commerce and Shopping."

*Variation in the GDPR's effects as a function of website popularity.* To examine the role of website popularity in GDPR's effect distribution, we split the websites into deciles according to their global, country, and industry ranks. While the country and industry ranks are initially reported separately for each country and industry that a website belongs to, we group the corresponding ranks for all countries and industries together, respectively, for the assignment into deciles. That way, the 10% most popular websites (i.e., the ones with the lowest rank numbers) worldwide, across all countries and industries, are part of the 1st decile, while the 10% least popular websites are part of the 10th decile. Figure 5 shows the distribution of the average effect of GDPR on the websites based on the industry rank deciles. Analyses based on the global and country ranks result in similar distributions.

Website popularity plays an important role in the effect distribution: Less-popular websites suffer from more negative effects than popular ones. Specifically, websites within the four bottom deciles exhibit far more negative effects than websites within other deciles. While the least popular websites (i.e., those in the bottom decile) suffer the most from GDPR (up to a 21% drop in total visits 18 months after the GDPR), websites within the 6th-9th industry-rank deciles exhibit a drop in the number of visits by, on average, 4.30%-6.23% after 3 months, and by 10.31%-11.51% after 18 months. Interestingly, the websites in the top decile (i.e., the most popular websites) show more negative effects, and even more so over time (3.74% after 3 months; 9.04% after 18 months), than websites in the 2nd-5th deciles (1.82%-2.96% after 3 months; 5.82%-7.25% after 18 months).

Still, the overall trend shows that users react less negatively to the changes induced by GDPR on more popular websites than the less popular ones, suggesting that the market is more concentrated after GDPR. This increase in market concentration is strongest in the first 3-9 months after GDPR but still exists 1.5 years after the GDPR. At the same time, research suggests that companies can employ several methods to mitigate the negative effect of privacy regulations on the competition by, e.g., reducing privacy concerns (Bleier et al. 2020; Turjeman and Feinberg 2020). Such mitigations might even result in companies experiencing a positive influence from privacy regulations. While we observe that there are websites that experience such positive effects, we also observe that, on average, GDPR affects websites negatively. Thus, the increased market concentration we observe post-GDPR suggests that websites should consider employing the methods discussed in the literature cited above to mitigate such negative effects.

*Variation in the GDPR's effects as a function of users' country of origin.* To examine the relationship between users' country of origin and the effect of GDPR, we categorized each website according to its most popular user base's country of origin (recall that our dataset provides user activity data corresponding to the country in which the website is most popular, as well as data corresponding to users in the US). Figure 6 suggests that the effects of the GDPR vary as a function of users' country of origin. Websites whose primary user base is from Denmark, Poland, or Germany suffered the least from GDPR over the analyzed period: 3 months after GDPR, the number of visits from users based in these countries decreased, on average, by 1%, 2.3%, and 2.9%, respectively. The strongest drops in website visits were associated with users from Austria, the Netherlands, UK, Hungary, Sweden, and Switzerland.

*Robustness of Results*

Our analysis required us to make several decisions, which may have impacted our findings. We categorize these decisions into five groups: website selection, control group, data, confounding factors, and SCG method. In Web Appendices C-H, we examine how strongly our decisions impact our results. Here, we summarize the analyses' main results.

*Website selection.* We selected a threshold of an average of 1,000 visits per week and removed all websites with fewer visits from our sample. To examine the sensitivity of the results for the chosen threshold, we conduct two robustness checks: 1) reduce the threshold to 700, and 2) increase the threshold to 2,000. As we show in Web Appendix C, our findings are robust to decreasing the threshold to 700 visits per week and increasing it to up to 1,700 visits per week – but not further. So, it does not look like our chosen threshold impacted our results.

*Figure 4: Distribution of the Effect of GDPR across Website Industries*

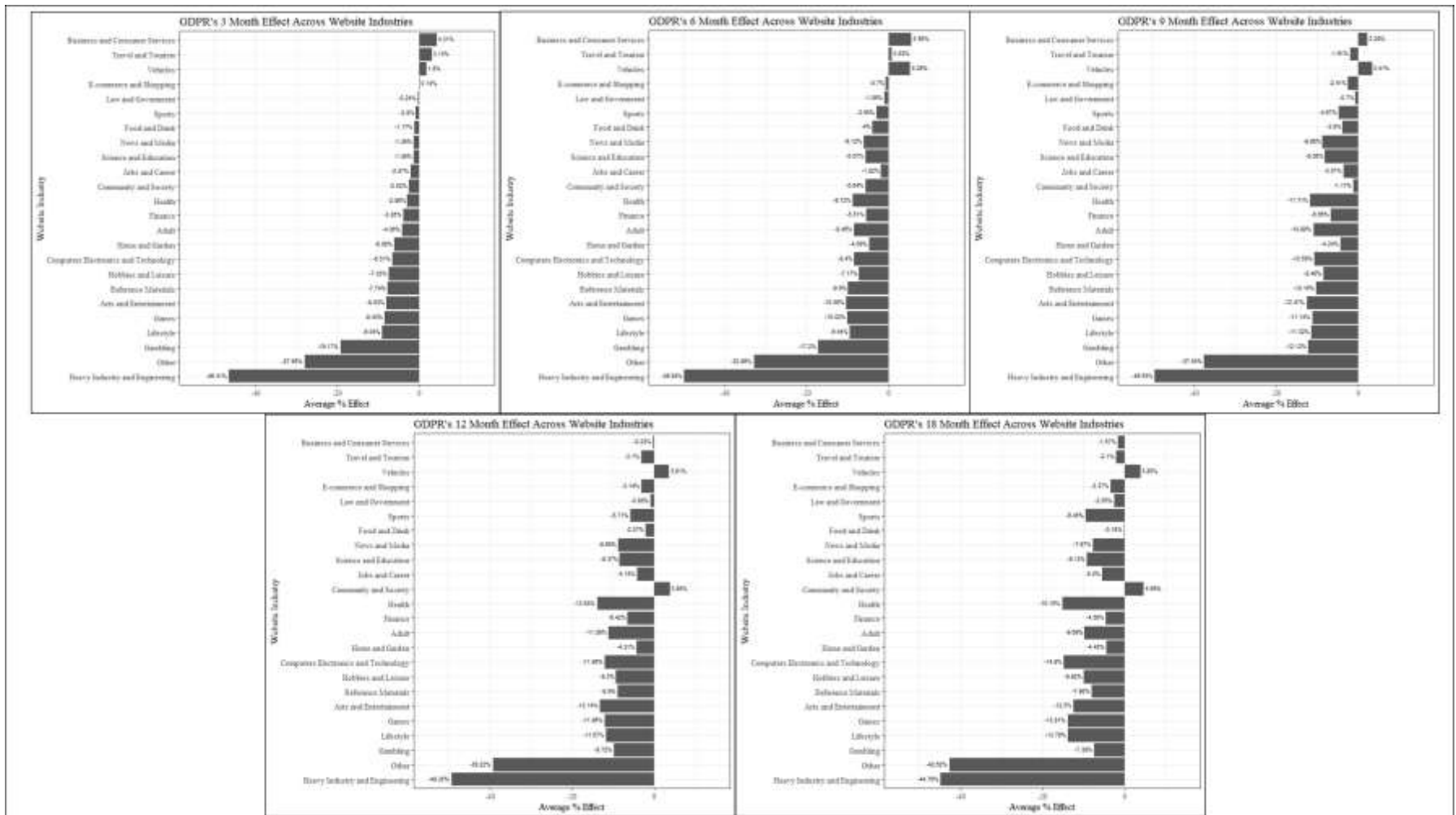

Reading example: The value of 4.31% in the upper left panel (i.e., the figure with the GDPR's 3-month effect across website industries) means that, on average, GDPR increases the total number of visits of the websites in the business and consumer services industry by 4.31%.

*Figure 5: Distribution of the Effect of GDPR across Deciles of Industry Ranks*

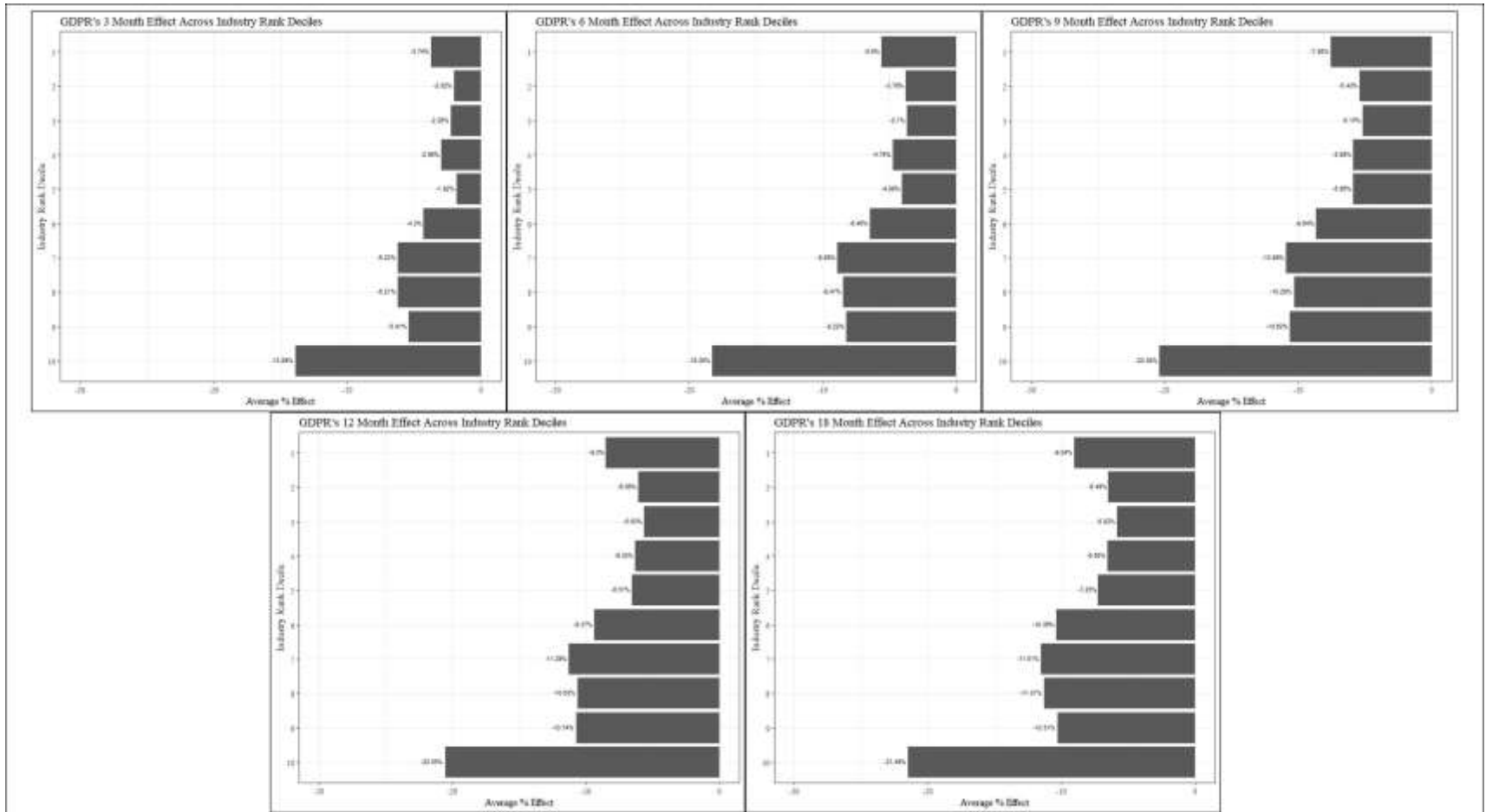

Reading example: The value of -3.74% in the upper left panel (i.e., the figure with the GDPR's 3-month effect across industry rank deciles) means that, on average, GDPR reduces the total number of visits across the most popular (i.e., those in the top-decile) websites in each industry by -3.74%. The results of the top-decile reflect the change of the 10% highest-ranked websites over the 24 industries.

*Figure 6: Distribution of GDPR Effect across User Countries*

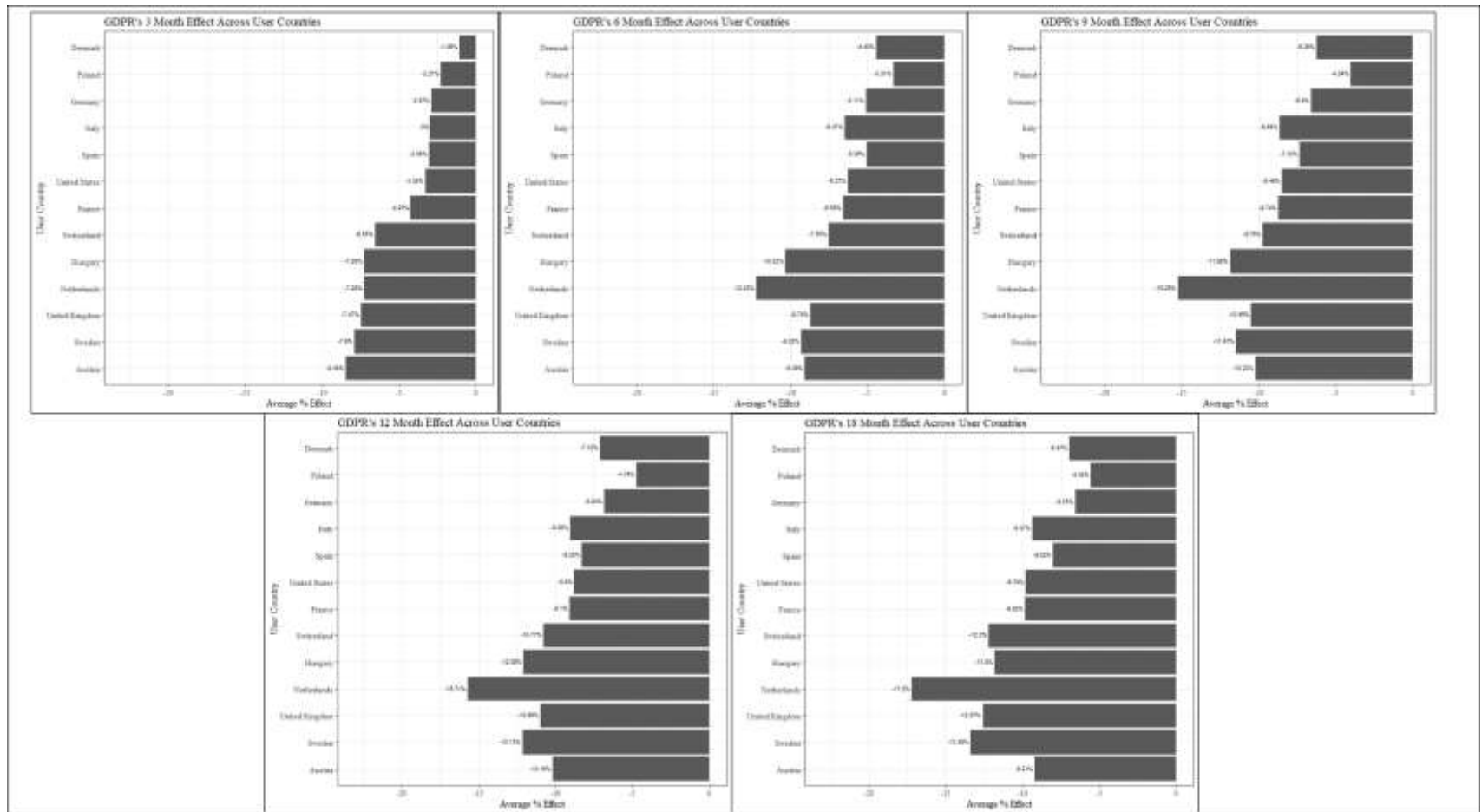

Reading example: The value of -1.06% in the upper left panel (i.e., the figure with the GDPR's 3-month effect across user countries) means that, on average, GDPR reduces the total number of visits of users coming from Denmark by 1.06%.

*Control group.* For our control group, we use website-instances of Non-EU-websites and Non-EU-users. Still, Non-EU-websites could voluntarily comply with GDPR for Non-EU-users, resulting in a potential spillover effect of GDPR to our control group (i.e., the so-called Brussels effect). Such a spillover effect would reduce the suitability of our control group as some control websites would be treated as well. This spillover effect of voluntary compliance could take two potential forms: First, websites could adapt their data storage systems to accommodate GDPR requirements for all users. Second, websites could adapt their user interface (e.g., privacy policy, consent banners) for all users. While rumors exist for the first form (e.g., Microsoft), they do not exist for the second form. Instead, websites show different user interfaces depending on a user's location (e.g., The Washington Post). As we measure the effect of websites implementing GDPR on their website, we only capture the second form of spillover effects in our analysis. Still, we further examine our control group.

The incentive for voluntary compliance is likely higher for Non-EU-websites if the share of EU-users is higher (i.e., websites comply with GDPR for a large share of users). Thus, we divide Non-EU-websites into deciles based on their EU-user-share and examine the results for our main metric. Web Appendix E shows that our main metric increases across all deciles for the control group. We further find no relation between the EU-user-share of control websites and the log-difference between the Non-EU-user base before and after the GDPR.

Furthermore, a spillover effect of GDPR to our control group would become apparent if the behavior of Non-EU-users on EU-websites (part of our treatment group) does not differ from their behavior on Non-EU-websites (control group). Similarly, a spillover effect would be likely if the behavior of EU-users on the Non-EU-websites (part of our treatment group) does not differ from the behavior of Non-EU-users on the same websites (control group). Web Appendix E shows that these differences in user behavior occur. Furthermore, a manual inspection of a subsample of control websites indicates no voluntary compliance. Overall, a spillover effect is not present or not large enough to change our results in size or directions.

Furthermore, websites based in the EU before GDPR could have decided to relocate to a Non-EU location to avoid having to comply with GDPR for Non-EU-users. These strategic shifts would reduce the suitability of our control group as some websites would have self-selected themselves to belong to the control group. Thus, we examine these possible strategic shifts performed by websites: We re-calculate the GDPR's effect on our main metric for a subsample with a stricter control group that includes only websites with domain suffixes indicating a Non-EU location. Web Appendix E shows no significant difference in the GDPR's effect for the two approaches.

*Data.* We utilize a dataset provided by SimilarWeb. Although companies (e.g., Google, Alibaba) and researchers (e.g., Calzada and Gill 2020, Lu et al. 2020) use SimilarWeb's data, SimilarWeb is not very transparent about its data collection procedures and whether GDPR affects SimilarWeb in its data collection methods. Therefore, we compare the quality of the SimilarWeb data for a subset of websites with another data source (German AGOF, https://www.agof.de/en). AGOF is a highly reliable and certified data source trusted by the German media market, making it the official traffic source in Germany. Additionally, it is very transparent in its data collection procedure. The analysis within Web Appendix F shows no significant difference in terms of the number of unique visitors and number of page impressions between those two data sources after GDPR. AGOF states that GDPR does not directly impact its metrics but that users might change their interaction with websites due to GDPR– which is what we want to measure in this paper.

*Confounding factors.* The enforcement of GDPR might have coincided with other changes in potentially confounding factors, such as an increase in internet speed that differs between our control and treatment groups and changes in user behavior on websites, affecting the estimated treatment effect. Therefore, in Web Appendix G, we investigate whether there have been substantial changes in confounding factors for Non-EU-users and EU-users (i.e., internet speed, the share of the population with internet access/laptop/smartphone). We find no evidence for a difference between the user groups in the pre- and post-GDPR comparison across the factors mentioned above that could substantially increase the control group's user behavior.

*Synthetic control method.* When calculating GDPR's effects in our analysis, we had to make several decisions regarding the method used, the requirement of the control and treated websites belonging to the same industry, and the number of control websites. As these decisions might impact our results, we examine the sensitivity of our estimates to the selected specifications.

Firstly, regarding the method used, we re-calculate the GDPR's effect for our main metric for a subsample using the original SCG method (Abadie et al. 2015; Bell et al. 2017), applied in the R package "synth" – instead of the R package "gsynth." Secondly, our SCG consisted of five websites within the same industry as our treated website with the highest pre-treatment correlations. To reveal the impact of these decisions, we repeated the analysis with the SCG consisting of 1) five websites with the highest pre-treatment correlations irrespectively of the industry (instead of belonging to the same industry), 2) five websites that are matched based on the EU-traffic share (instead of belonging to the same industry), and 3) ten websites (instead of 5). Web Appendix H shows no significant difference in the observed effects, indicating that our results are robust to these specifications.

*DISCUSSION*

*Summary of Results: User Quantity and Usage Intensity*

The results of our user quantity and usage intensity analyses show that, on average, and at each time point investigated, the GDPR negatively affected most user quantity and usage intensity metrics (except for the page impressions per visit). We further observe that the negative effects of GDPR become stronger over time: 3 months after the GDPR, the user quantity metrics of treated website-instances drop on average by 0.8%-4.9%; 18 months after the GDPR, the average values of these metrics are 6.6%-10.2% below their pre-GDPR levels. These findings highlight the importance of investigating the effects of the GDPR over a longer period. Overall, though GDPR affected some websites positively, it affected most (62%-67%, depending on the user quantity metric) of the websites negatively after 18 months.

As outlined in Table 3, our user quantity metrics depend on one another: All user quantity metrics are a function of the total number of visits and one corresponding usage intensity metric. Thus, it is not too surprising that the directions of the effects of the user quantity metrics are aligned. The predominant difference between the GDPR's effect on the user quantity metrics is the effect sizes – differences driven by the usage intensity metrics.

Concerning usage intensity, we observe that the average effect of GDPR on usage intensity is generally negative in the first 3 months of GDPR (i.e., the bounce rate rises by 0.8%, the number of visits per unique visitor, and the time spent per website visit decrease by 1.6% and 0.8%, respectively; only the page impressions per visit show a positive effect with a 2% increase), but the negative effects become less strong over time. After 18 months of GDPR, the number of visits per unique visitor is only 0.6% below the pre-GDPR baseline, and the number of page views and the time spent per visit even increase by 2.2% and 0.1%, respectively. Only the bounce rate shows a negative trend—it increases by 2.5%.

The effect distribution across websites reveals that the GDPR's effects on usage intensity metrics partly balance out the effects on user quantity, i.e., an increase in user quantity usually goes along with a decrease in usage intensity and vice versa. For example, among websites that lose unique visitors after GDPR, the remaining visitors use the website more intensively than they did pre-GDPR: The average user generates more visits to those websites (e.g., 4.8% more visits per unique visitor 18 months post-GDPR), and engages more with the websites in each visit, as reflected in increases in the number of page impressions (+4.6%) and the time spent per visit (+2.1%). However, the user base exhibits an increased bounce rate, although the absolute number of bouncing visitors decreases.

Websites that gain unique visitors experience the opposite effect: The number of visits per unique visitor is lower post-GDPR than pre-GDPR (e.g., 9% lower at 18 months post-GDPR), and these unique visitors spend less time on each visit (-4%) and view fewer pages per visit (-4.6%), as compared with pre-GDPR unique visitors. Together with the increasing number of visits, the number of bouncing visitors rises, although the bounce rate decreases slightly.

Together, these results suggest that the GDPR negatively affects the average website in one of two major ways: Either the website experiences difficulties in attracting users, or, having attracted users, it struggles to keep them engaged and get them to return.

Existing literature shows similar effects of privacy regulations. For example, Goldberg et al. (2021) show that GDPR reduces the page impressions by about 12%, on average, for affected websites in the first 6 months of GDPR. We observe a weaker but still strong negative effect of GDPR.

Furthermore, we discussed the potential reasons for the positive and negative effects in our literature review and show that GDPR affects websites differently: For some websites, the user base shows positive effects due to the website's adjustment to GDPR, potentially due to lower privacy concerns (e.g., Martin 2015) or higher trust in the website (e.g., Martin et al. 2017). The user base reacts negatively for other websites, potentially due to the new awareness of data disclosure activities or increased privacy concerns (e.g., Dinev and Hart 2006). Again, the user base does not change the behavior for other websites, potentially due to the actual behavior not reflecting the stated privacy preferences (e.g., Acquisti 2004) or a continued feeling of powerlessness (Few 2018) even after the enforcement of GDPR.

The diverse reactions from users shown in the literature and observed in this study show the complex relationship between the user's attitudes towards privacy and the resulting user behavior. Following the GDPR, users might not only change their website visiting behavior, but also their reaction to online advertising or targeting, or their general attitudes towards websites – especially those that react negatively to the GDPR adjustments by websites (as discussed by Kim et al. 2018 in the case of websites' transparency regarding data disclosure).

*Differential Effects of the GDPR as a Function of Website and User Characteristics*

Our results show that that the effects of the GDPR varied across different websites. In particular, less popular websites were hurt the most: 18 months post-GDPR, the 10% least popular websites experienced an average drop of 21% in the total number of visits. The 10% most popular websites, in contrast, experienced an average drop of only 9% in total visits. These results suggest that the GDPR increased market concentration. These effects may reflect users' stronger motivation to continue using more popular and valued websites despite any potential disadvantages created by the GDPR—including, for example, users' heightened awareness of data disclosure or diminished convenience of use due to website compliance.

For less popular websites, users may be less likely to feel that the benefits of continued use outweigh the disadvantages.

We further observed that the effects of the GDPR varied across websites from different industries. For example, the most negatively affected websites included those in the Entertainment and Lifestyle segment (7.4%-13.8% decrease in visits after 18 months of GDPR). Other types of websites experienced positive effects (e.g., Vehicles with an increase of 3.9% in total visits after 18 months). These effect differences may indicate differences in users' expectations regarding privacy across industries. For example, users visiting entertainment websites may previously have been less aware of data collection than users on, e.g., e-commerce websites, where it is necessary to provide information to purchase products. Consequently, highlighting data collection practices may have been more "surprising" to users of entertainment websites and had a stronger effect on their behavior. Likewise, as in the case of more popular websites, users seeking services in domains that they deem more necessary may feel that the advantages of continuing to use the website outweigh the disadvantages—and in some cases, they may even value the added safeguards on their privacy.

Finally, we observed that the GDPR's effects differed across users from different countries of origin, reflecting cultural differences across countries. For example, users from Denmark, Poland, Germany, Italy, and Spain reacted less negatively to the GDPR than users from the Netherlands, Sweden, and the UK.

## ANALYSIS OF GDPR'S ECONOMIC EFFECT ON WEBSITES

Our study focused on quantifying the effect of the GDPR on user quantity and usage intensity. However, what is also of interest in policymakers' tradeoffs is the extent to which the GDPR damages companies' revenue, as this damage is likely to cause negative societal effects. In what follows, we outline a back-of-the-envelope estimation of the magnitude of possible economic effects on the websites resulting from changes in user behavior due to GDPR. In these estimations, we rely on the average effects of GDPR after 18 months as the basis for our calculations. We present two different estimations corresponding to two kinds of websites: 1) websites that earn money by selling products, i.e., e-commerce websites, and 2) websites that earn money by displaying ads.

*Analysis of GDPR's Economic Effect on E-Commerce Websites*

For the e-commerce websites within this study's sample, the average drop in total visits at 18 months post-GDPR amounted to 3.37% (see Figure 4). The determining revenue factors of an e-commerce website $w$ are the number of visits (i.e., non-unique visitors), the conversion rate (i.e., the share of visits resulting in a purchase), and the revenue per purchase:

(6) $Revenue = Number\ of\ Visits * Conversion\ Rate * Revenue\ per\ Purchase$

Based on the Q1 2020 e-commerce benchmarks by Monetate (2020), the revenue per purchase globally is $105.99, and the average conversion rate per visit is 1.91%. Looking at the e-commerce websites within the website sample of this study, the average yearly total number of visits across all countries amounted to 70,461,862. Thus, the average yearly revenue for an e-commerce website within this study's sample before GDPR amounts to:

(7) $Revenue_{avg.} = 70,461,862 * 1.91\% * \$105.99 = \$142,643,623.54.$

The average drop in total visits 18 months (=1.5 years) after the GDPR to the e-commerce websites within the study's sample represents the respective drop in website visits:

(8) $Revenue\ Change_{18\ months} = -3.37\% * \$142.643.623,54 * 1.5 = -\$7,209,722.73.$

A decrease of 3.37% in total visits due to GDPR can thus decrease the revenue of an average e-commerce website by over $7 million in the first 18 months after enforcement of the GDPR.

*Analysis of GDPR's Economic Effect on Ad-Based Websites*

For an ad-based website *w*, the determining factors for the revenue are the number of page impressions, the number of ads displayed per page impression, and the price per ad impression.

(9) $Revenue = No. of\ Page\ Impressions * Ads\ per\ Page\ Impression * Ad\ Price$

As an example of an ad-based industry, we examine the economic effect of the GDPR on websites within the News and Media industry. In our sample, the average number of yearly page impressions on a news and media website across all regions was 358,859,344. A random sample of the homepages and article pages of 7 important news websites (nytimes.com, huffpost.com, washingtonpost.com, news.yahoo.com, bbc.com, wsj.com, and cnn.com) shows an average of 7.6 ads per page. Based on both ComScore (2010) and TheBrandOwner (2017), the average CPM (cost for a thousand ad impressions) for news websites lies between $7 and $8 (here $0.0075 per Impression). Using these values, the total yearly revenue for an average news website before the GDPR for the analyzed website sample amounts to:

(10) $Revenue_{avg.} = 358,859,344 * 7.6 * \$0.0075 = \$20,454,982.61.$

The average effect of GDPR on page impressions after 18 months on our sample of news and media websites is a drop of 8.05%. This reduction in the number of page impressions decreases the revenue of news websites significantly:

(11) $Revenue\ Change_{18\ months} = -8.05\% * \$20,454,982.61 * 1.5 = -\$2,469,953.43.$

A decrease in page impressions of 8% due to the GDPR can thus decrease the revenue of an average news and media website by almost $2.5 million in the first 1.5 years after the GDPR.

*CONCLUDING REMARKS*

Our analysis reveals that, during the 1.5 years following the GDPR's enforcement, the law's affected user quantity and usage intensity (among the websites to which the privacy law applied) negative, on average. Furthermore, the effects increase over time, highlighting the importance of examining the effects of the privacy laws' enforcement over time as it might take some time for their full effect to become apparent. Our results further suggest that the effects were not evenly distributed across websites, with less-popular websites and websites within certain industries being more strongly affected than others—and in fact, some websites were positively affected. We have shown how to use our findings to assess the economic loss that websites might suffer due to the enforcement of new privacy laws. Our results rest upon the assumption that our control group is unaffected by GDPR. We provide support for this assumption but cannot entirely exclude that the treatment also impacted the control group. If that were the case, the actual treatment effect would be even larger than our estimated treatment effect so that our results would represent a lower bound of GDPR's effect.

Our results provide policymakers with an assessment of how the GDPR becoming effective has changed users' interactions with websites, the financial outcomes of websites, and the competitive landscape. This information is of interest in itself and can be used to assist policymakers and companies in making inferences and predictions about similar upcoming privacy laws. In particular, policymakers might use our results to adjust privacy laws in the drafting stage or issue guidelines and frameworks complementing already enforced or approved privacy laws. In this regard, we note that, given the novelty of the GDPR (as the first major European privacy law since the e-Privacy Directive in 2002), coupled with its strict nature, users' reactions to the GDPR might potentially be stronger than their reactions to other, less strict privacy laws. Thus, the effect sizes found in this paper may represent an upper bound for the potential effects of other privacy laws.

It is important to acknowledge that underlying our results is a complex pattern of user behavior, reflecting users' responses to changes in websites' privacy policies and consent banners (e.g., requests for consent to data collection), influenced by a calculus of the benefits and disadvantages of continuing to use particular websites, given these new policies. Yet, it seems reasonable to assume that overall use of the internet is unlikely to change due to a privacy law such as GDPR. Consequently, users might reallocate their usage to spend more time on the websites they value or trust.

Indeed, our results are compatible with this type of reallocation. The sunk-cost effect could explain this change in usage allocation across websites (Arkes and Blumer 1985), which states that 1) users tend to use a product or service more if they know what it cost them, and 2) users want to use paid-for services more to not feel like they wasted their money. In this scenario, a privacy law may make users aware of the "cost" of a website's services—i.e., users pay with their data—leading users to prefer websites to which they have already provided data. These underlying patterns of user behavior are outside the scope of the current research. Still, the empirical exploration of the mechanisms underlying our observed behavioral trends opens up further research opportunities.

# The Impact of Privacy Laws on Online User Behavior

# Web Appendix

## Table of Contents of Web Appendix

*WEB APPENDIX A: ADDITIONAL FIGURES*

Figure W 1 shows the number of websites for the different industries within our final sample. The sample includes websites that span diverse industries, with the industries such as "Computers, Electronics and Technology," "News and Media," and "Arts and Entertainment" being the most represented.

*Figure W 1: Distribution of Websites across Industries*

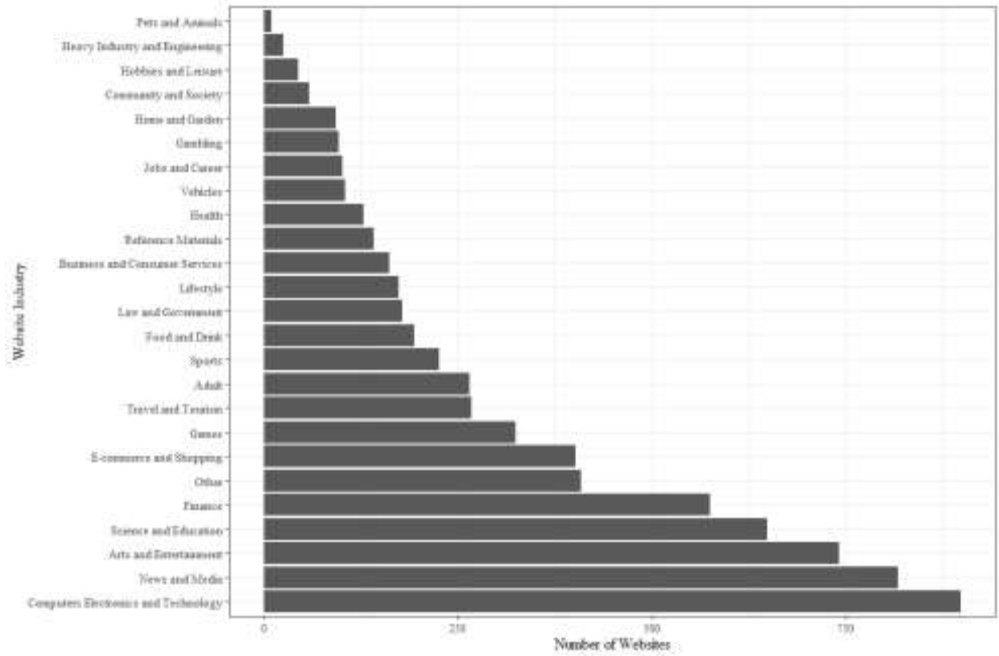

Figure W 2 shows the distribution of GDPR's effect for the websites' number of unique visitors over the different post-treatment periods. The solid line plot indicates the share of websites that GDPR affects negatively, and the dashed line plot the share of websites that GDPR affects significantly (on the 5%-level).

*Figure W 2: Distribution of the Effect of GDPR on Monthly Number of Unique Visitors over Time*

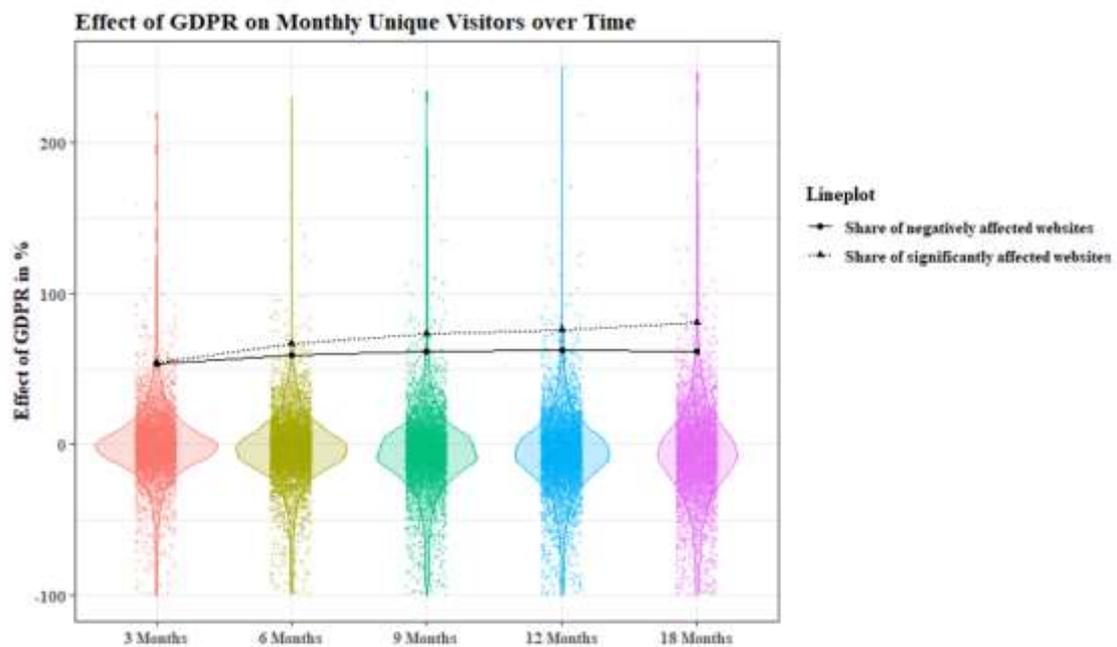

Figure W 3 shows the distribution of GDPR's effect for the websites' number of page impressions over the different post-treatment periods. The solid line plot indicates the share of websites that GDPR affects negatively, and the dashed line plot the share of websites that GDPR affects significantly (on the 5%-level).

*Figure W 3: Distribution of the Effect of GDPR on Weekly Number of Page Impressions over Time*

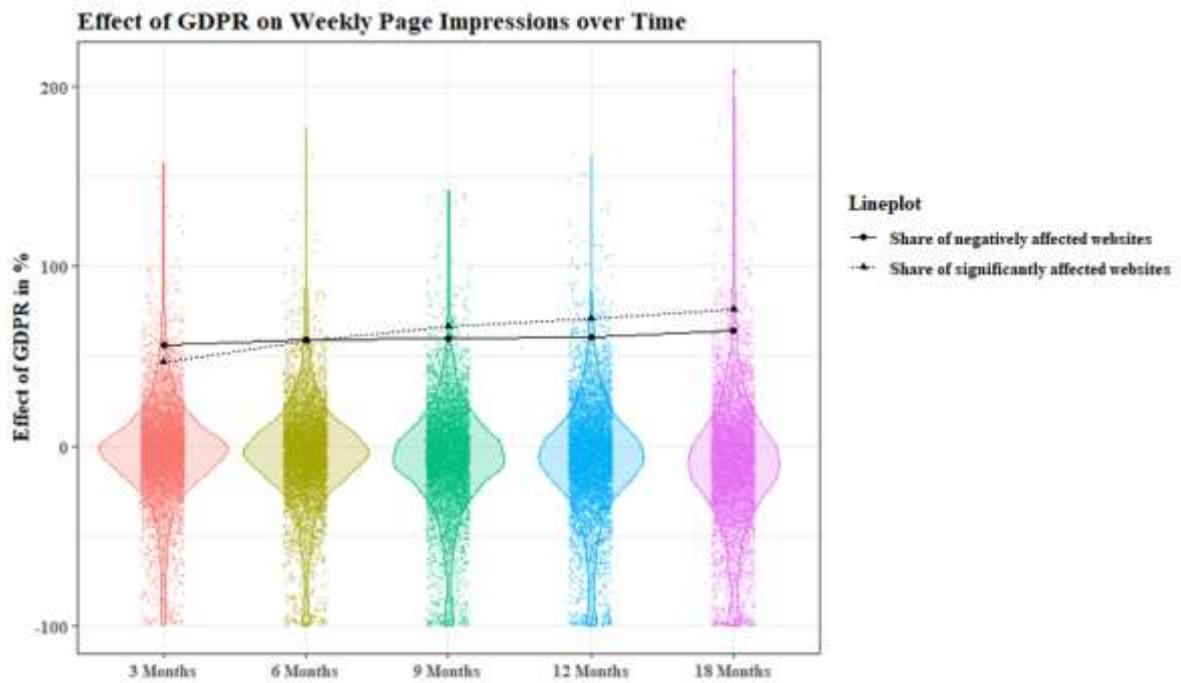

Figure W 4 shows the distribution of GDPR's effect for the weekly time on the websites over the different post-treatment periods. The solid line plot indicates the share of websites that GDPR affects negatively, and the dashed line plot the share of websites that GDPR affects significantly (on the 5%-level).

*Figure W 4: Distribution of the Effect of GDPR on Weekly Time on Website over Time*

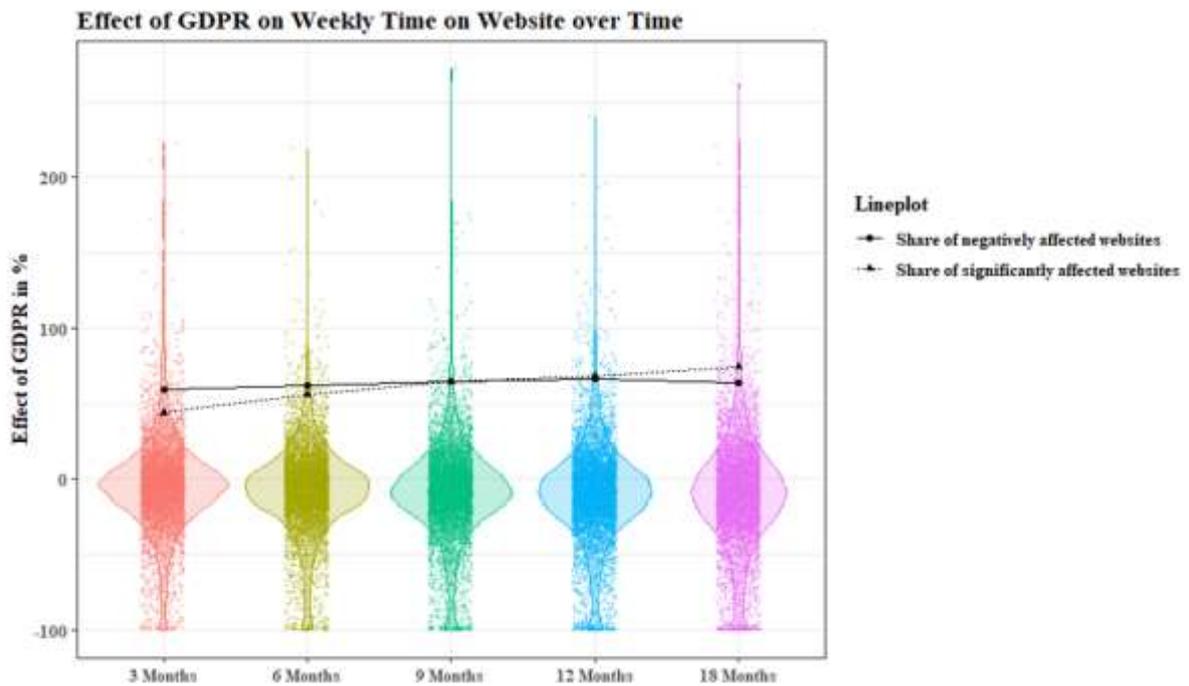

Figure W 5 shows the distribution of GDPR's effect on the websites' number of bouncing visitors over the different post-treatment periods. The solid line plot indicates the share of websites that GDPR affects negatively, and the dashed line plot the share of websites that GDPR affects significantly (on the 5%-level).

*Figure W 5: Distribution of the Effect of GDPR on Weekly Number of Bouncing Visitors over Time*

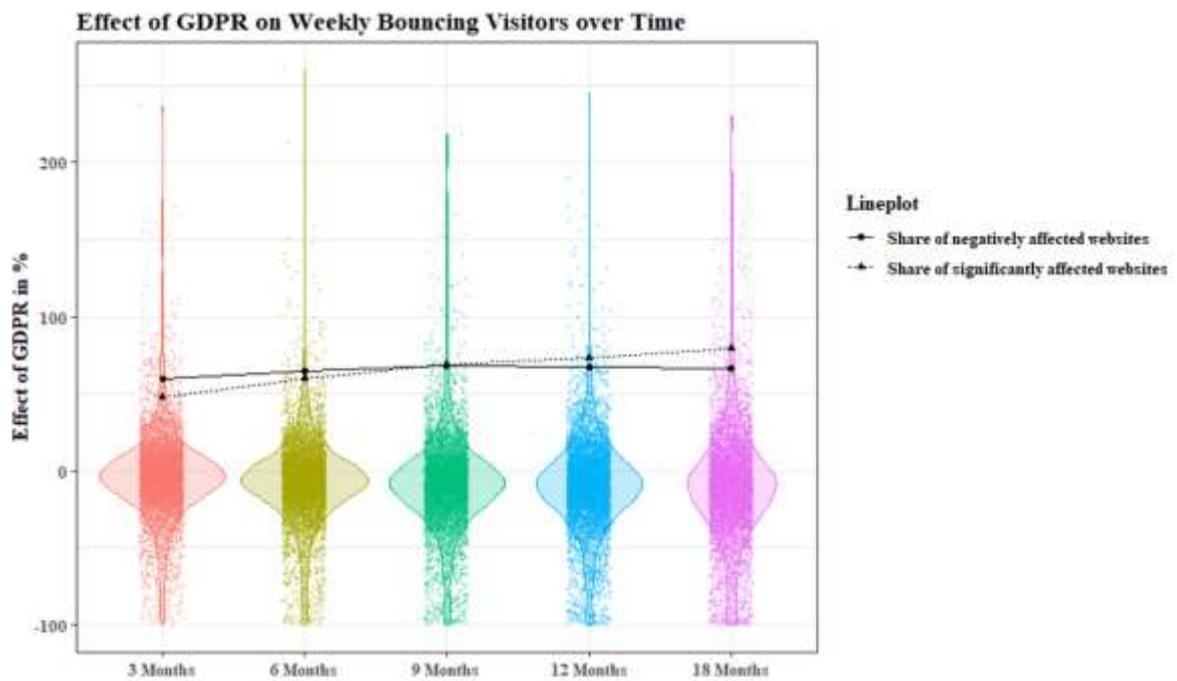

*WEB APPENDIX B: DERIVATION OF ANALYSIS OF USAGE INTENSITY METRICS*

Every usage intensity metric is a function of two user quantity metrics, as shown in Table 3. For example, we calculate the number of visits per unique visitor on a website *w* by dividing the number of unique visitors on the website *w* by the total number of visits on website *w*; the other usage intensity metrics follow the same logic (see Table 3):

(W1) $Number\ of\ Visits\ per\ Unique\ Visitor_w = \frac{Total\ Number\ of\ Visits_w}{Number\ of\ Unique\ Visitors_w}$

(W2) $Page\ Impressions\ per\ Visit_w = \frac{Number\ of\ Page\ Impressions_w}{Total\ Number\ of\ Visits_w}$

(W3) $Time\ per\ Visit_w = \frac{Time\ on\ Website_w}{Total\ Number\ of\ Visits_w}$

(W4) $Bounce\ Rate_w = \frac{Number\ of\ Bouncing\ Visitors_w}{Total\ Number\ of\ Visits_w}$

We use Equations (W1)-(W4) to examine the GDPR's effect on the usage intensity metrics in the following manner:

1) We use Equation (1) to calculate the GDPR's effect ($\Delta$) for all user quantity metrics. Our methodology to derive this effect $\Delta$ capturing the relative changes of each user quantity metrics for each website *w* and each post-treatment period *p*.

2) We then incorporate the GDPR's effect $\Delta$ for all the metrics for each post-treatment period *p* in Equations (W1)-(W4). As we capture the relative changes for the metrics, we can incorporate the GDPR's effect $\Delta$ for each post-treatment period *p* for each website *w* by multiplying the metrics' pre-treatment values with ($1+\Delta_{p,w}$).

3) Finally, we rearrange the equations to isolate the GDPR's effect $\Delta$ on the usage intensity metric of interest.

We now use the described process to derive the formula for the GDPR's effect on the usage intensity metric "bounce rate". The other usage intensity metrics follow the same logic and can be derived accordingly.

GDPR's effect $\Delta$ for post-treatment period $p$ for website $w$ for all metrics within Equation (W4):

(W5)  $Number\ of\ Bouncing\ Visitors_{p,w} = Number\ of\ Bouncing\ Visitors_{w} * (1 + \Delta Number\ of\ Bouncing\ Visitors_{p,w})$

(W6)  $Total\ Number\ of\ Visits_{p,w} = Total\ Number\ of\ Visits_{w} * (1 + \Delta Total\ Number\ of\ Visits_{p,w})$

(W7)  $Bounce\ Rate_{p,w} = Bounce\ Rate_{w} * (1 + \Delta Bounce\ Rate_{p,w})$

Incorporation of GDPR's effect $\Delta$ (Equations W5-7) into Equation (W4):

(W8)  $Bounce\ Rate_{p,w} = \frac{Number\ of\ Bouncing\ Visitors_{p,w}}{Total\ Number\ of\ Visits_{p,w}}$

(W9)  $Bounce\ Rate_{w} * (1 + \Delta Bounce\ Rate_{p,w}) = \frac{Number\ of\ Bouncing\ Visitors_{w} * (1 + \Delta Number\ of\ Bouncing\ Visitors_{p,w})}{Total\ Number\ of\ Visits_{w} * (1 + \Delta Total\ Number\ of\ Visits_{p,w})}$

Rearranging Equation (W5):

(W10)  $1 + \Delta Bounce\ Rate_{p,w} = \frac{Number\ of\ Bouncing\ Visitors_{w} * (1 + \Delta Number\ of\ Bouncing\ Visitors_{p,w})}{Total\ Number\ of\ Visits_{w} * (1 + \Delta Total\ Number\ of\ Visits_{p,w}) * Bounce\ Rate_{w}}$

(W11)  $\Delta Bounce\ Rate_{p,w} = \frac{Number\ of\ Bouncing\ Visitors_{w} * (1 + \Delta Number\ of\ Bouncing\ Visitors_{p,w})}{Total\ Number\ of\ Visits_{w} * (1 + \Delta Total\ Number\ of\ Visits_{p,w}) * Bounce\ Rate_{w}} - 1$

(W12)  $\Delta Bounce\ Rate_{p,w} = \frac{Number\ of\ Bouncing\ Visitors_{w}}{Total\ Number\ of\ Visits_{w}} * \frac{1}{Bounce\ Rate_{w}} * \frac{1 + \Delta Number\ of\ Bouncing\ Visitors_{p,w}}{1 + \Delta Total\ Number\ of\ Visits_{p,w}} - 1$

Inserting Equation (W4) into Equation (W8):

(W13)  $\Delta Bounce\ Rate_{p,w} = Bounce\ Rate_{w} * \frac{1}{Bounce\ Rate_{w}} * \frac{1 + \Delta Number\ of\ Bouncing\ Visitors_{p,w}}{1 + \Delta Total\ Number\ of\ Visits_{p,w}} - 1$

Final rearranging of Equation (W9):

(W14)  $\Delta Bounce\ Rate_{p,w} = \frac{1 + \Delta Number\ of\ Bouncing\ Visitors_{p,w}}{1 + \Delta Total\ Number\ of\ Visits_{p,w}} - 1$

In Equation (W10), we have isolated the GDPR's effect on the bounce rate and can examine this effect using the observed effects on the corresponding user quantity metrics, i.e., the number of bouncing visitors and the total number of visits. We can use Equation (W10) for the GDPR's effect on the bounce rate – and, in the same manner, for all usage intensity metrics – for each website and each post-treatment period $p$ (i.e., a period that covers 3, 6, 9, 12, and 18 months after the enforcement of GDPR).

*WEB APPENDIX C: ROBUSTNESS CHECKS*
*WITH RESPECT TO WEBSITE SELECTION*

In our analysis, we selected a threshold of an average of 1,000 visits per week for our sample. So, we removed all websites that had an average number of visits per week below this threshold. The chosen threshold might impact our results. Thus, we examine the sensitivity of the results for the chosen threshold with two robustness checks.

1) We decrease the threshold of visits and examine whether the findings of this study change significantly. More specifically, we decrease the threshold in steps of 100 visits per week until we reduce our initial threshold to 700 visits per week.

2) We increase the threshold of visits and examine whether the findings of this study change significantly. More specifically, we increase the threshold in steps of 100 visits per week until we double our initial threshold, i.e., we reach 2,000 visits per week.

*Decreasing the Threshold for Website Filtering*

Table W 1 shows how many additional websites we include when decreasing the filtering threshold. We further examine whether the inclusion of the additional websites results in the distribution of the obtained data sample's total visits being significantly different from the distribution of the original data sample's total visits. We find that reducing the threshold to 700 leads to the inclusion of 27 additional website-instances with EU-user data and 405 additional website-instances with Non-EU-user data. The additional website-instances, however, do not significantly affect the composition of websites within our data sample. Consequently, our findings are robust to decreasing the threshold, and it does not look like our chosen threshold eliminates too many websites with low traffic.

*Table W 1: Effect of Threshold Reduction on Number of Websites and Significant Difference between Original and New Sample*

| Threshold | Websites with EU-user data | | | | Websites with Non-EU-user data | | | |
|---|---|---|---|---|---|---|---|---|
| | **No. of additional Websites** | **Total Visits: Mean** | **Total Visits: Std. Deviation** | **p-value of t-test** | **No. of additional Websites** | **Total Visits: Mean** | **Total Visits: Std. Deviation** | **p-value of t-test** |
| 1,000 | +/-0 | 1,706,008 | 15,804,678 | - | +/-0 | 2,560,706 | 54,603,917 | - |
| | *Original sample* | | | | *Original sample* | | | |
| 900 | +10 | 1,703,242 | 15,791,435 | 0.95 | +110 | 2,509,291 | 54,052,980 | 0.74 |
| | *Original vs. new sample: No significant difference* | | | | *Original vs. new sample: No significant difference* | | | |
| 800 | +17 | 1,701,254 | 15,782,190 | 0.91 | +245 | 2,448,846 | 53,399,093 | 0.46 |
| | *Original vs. new sample: No significant difference* | | | | *Original vs. new sample: No significant difference* | | | |
| 700 | +27 | 1,698,450 | 15,769,008 | 0.86 | +405 | 2,381,041 | 52,654,007 | 0.23 |
| | *Original vs. new sample: No significant difference* | | | | *Original vs. new sample: No significant difference* | | | |

Notes: Significance level based on t-tests comparing the original sample with the newly obtained sample. Original sample refers to the sample of the main analysis with a threshold of 1,000 visits per week, as shown in Table 1.

*Increasing the Threshold for Website Filtering*

Table 7 shows how many websites we remove when increasing the filtering threshold. We further examine whether removing the additional websites yields a distribution of the obtained data sample's total visits that is significantly different from the distribution of the original data sample's total visits. We find that increasing the threshold to 1,700 results in removing 40 website-instances with EU-user data and 550 additional website-instances with Non-EU-user data. Increasing the threshold up to 1,700 does not significantly affect the composition of websites within our data sample. However, increasing the threshold to 1,800 or higher would significantly affect the composition of websites within our data sample. Thus, our findings are robust to increasing the threshold to 1,700. While a substantially higher threshold of 1,800 or higher results in removing too many websites with low traffic, it does not look like our chosen threshold eliminated too many websites with low traffic.

*Table W 2: Effect of Threshold Increase on Number of Websites and Significant Difference between Original and New Sample*

| Threshold | Websites with EU-user data | | | | Websites with Non-EU-user data | | | |
|---|---|---|---|---|---|---|---|---|
| | No. of additional Websites | Total Visits: Mean | Total Visits: Std. Deviation | p-value of t-test | No. of additional Websites | Total Visits: Mean | Total Visits: Std. Deviation | p-value of t-test |
| 1,000 | +/-0 | 1,706,008 | 15,804,678 | - | +/-0 | 2,560,706 | 54,603,917 | - |
| | *Original sample* | | | | *Original sample* | | | |
| 1,100 | -3 | 1,706,811 | 15,808,661 | 0.98 | -98 | 2,608,369 | 55,109,208 | 0.76 |
| | *Original vs. new sample: No significant difference* | | | | *Original vs. new sample: No significant difference* | | | |
| 1,200 | -11 | 1,709,014 | 15,819,288 | 0.94 | -197 | 2,658,303 | 55,634,171 | 0.53 |
| | *Original vs. new sample: No significant difference* | | | | *Original vs. new sample: No significant difference* | | | |
| 1,300 | -13 | 1,709,594 | 15,821,946 | 0.93 | -282 | 2,702,767 | 56,097,071 | 0.37 |
| | *Original vs. new sample: No significant difference* | | | | *Original vs. new sample: No significant difference* | | | |
| 1,400 | -21 | 1,711,915 | 15,832,591 | 0.89 | -363 | 2,746,250 | 56, 549,116 | 0.24 |
| | *Original vs. new sample: No significant difference* | | | | *Original vs. new sample: No significant difference* | | | |
| 1,500 | -28 | 1, 713,875 | 15,841,931 | 0.85 | -428 | 2,782,440 | 56,919,877 | 0.17 |
| | *Original vs. new sample: No significant difference* | | | | *Original vs. new sample: No significant difference* | | | |
| 1,600 | -35 | 1,715,880 | 15,851,283 | 0.82 | -490 | 2,817,770 | 57,280,410 | 0.11 |
| | *Original vs. new sample: No significant difference* | | | | *Original vs. new sample: No significant difference* | | | |
| 1,700 | -40 | 1,717,139 | 15,857,986 | 0.79 | -550 | 2,852,213 | 57,635,926 | 0.07 |
| | *Original vs. new sample: No significant difference* | | | | *Original vs. new sample: No significant difference* | | | |
| 1,800 | -46 | 1,718,852 | 15,866,027 | 0.76 | -621 | 2,894,595 | 58,065,293 | 0.04 |
| | *Original vs. new sample: No significant difference* | | | | *Original vs. new sample: Significant difference* | | | |
| 1,900 | -53 | 1,720,868 | 15,875,422 | 0.73 | -677 | 2,929,203 | 58,410,794 | 0.03 |
| | *Original vs. new sample: No significant difference* | | | | *Original vs. new sample: Significant difference* | | | |
| 2,000 | -57 | 1,722,026 | 15,880,798 | 0.71 | -732 | 2,964,009 | 58,756,199 | 0.02 |
| | *Original vs. new sample: No significant difference* | | | | *Original vs. new sample: Significant difference* | | | |

Notes: Significance level based on t-tests comparing the original sample with the newly obtained sample. Original sample refers to the sample of the main analysis with a threshold of 1,000 visits per week, as shown in Table 1.

## *WEB APPENDIX D: ROBUSTNESS CHECKS*
## *WITH RESPECT TO EARLY AND LATE COMPLIANCE*

We use the enforcement date of GDPR (May 25th, 2018) to construct a before-and-after analysis, comparing the treatment group to the control group to quantify the intention-to-treat effect of GDPR. Although very few websites were compliant before the enforcement date (Hochstadt 2018), we examine whether websites were possible early or late with their compliance with GDPR. Such early or late compliance might affect the validity of our results as the timing of the treatment effect would differ from the GDPR's enforcement date.

In our robustness check, we perform the same calculations as we describe in our methodology section. The only difference is that we did not use the entire observation period for the analysis but removed the observations 30 days before the enforcement of GDPR and 30 days after. We then calculate the GDPR's short- and long-term effect on our user quantity metrics for all websites. We observe no significant differences between the GDPR's effect across the websites for the short- or long-term for all our user quantity metrics. We show the results of the robustness check for our main metric, the total number of visits, in Table W 3. Thus, there is no cause for concern that early or late compliance of websites might influence our findings.

*Table W 3: Summary of Results for Total Visits with and without Inclusion of 30-Day-Period before and after GDPR*

| Total Visits | Original Results (entire observation period) | | Robustness Results (omitting 30 days before and after enforcement date) | |
|---|---|---|---|---|
| | **Median** | **Mean** | **Median** | **Mean** |
| **3 months** | -3.49% | -4.88% | -4.55% | -5.49% |
| | **p-value of t-test:** 0.26 *No significant difference between original and robustness results* | | | |
| **6 months** | -5.54% | -7.22% | -6.38% | -7.91% |
| | **p-value of t-test:** 0.17 *No significant difference between original and robustness results* | | | |
| **9 months** | -7.54% | -9.07% | -8.24% | -9.74% |
| | **p-value of t-test:** 0.20 *No significant difference between original and robustness results* | | | |
| **12 months** | -8.24% | -9.57% | -8.78% | -10.09% |
| | **p-value of t-test:** 0.35 *No significant difference between original and robustness results* | | | |
| **18 months** | -8.91% | -10.02% | -9.56% | -10.30% |
| | **p-value of t-test:** 0.65 *No significant difference between original and robustness results* | | | |

*WEB APPENDIX E: ROBUSTNESS CHECKS WITH RESPECT TO CONTROL GROUP*

For the control group specification, we use website-instances of Non-EU-websites and Non-EU-users. However, there might be a concern about a potential spillover effect of GDPR to website-instances that are not within the applicable scope of the privacy law. For instance, Non-EU-websites with visitors from EU locations have to comply with GDPR for those visitors. It could be that differentiating between EU- and Non-EU-users is more costly than simply complying with GDPR for all users, i.e., even for Non-EU-users. As a result, a website might voluntarily treat Non-EU-users with GDPR compliance (i.e., the so-called Brussels effect). This spillover effect of voluntary compliance could take two potential forms: First, websites could adapt their data storage systems to accommodate GDPR requirements for all users. Second, websites could adapt their user interface (e.g., privacy policy, consent banners) for all users. While rumors exist for the first form (e.g., Microsoft), they do not exist for the second form. While we would only capture the second form in our analysis, such a spillover effect might influence the adequacy of our control group as some websites within the control group would be treated.

The incentive for voluntary compliance is likely higher for Non-EU-websites if the share of EU-users is higher (i.e., websites comply with GDPR for a large share of users). A potential spillover effect of GDPR to our control group will also become apparent if the behavior of Non-EU-users on EU-websites (part of our treatment group) does not differ from their behavior on Non-EU-websites (control group) and if the behavior of EU-users on the Non-EU-websites (part of our treatment group) does not differ to the behavior of Non-EU-users on the same websites (control group).

Thus, to examine whether potential spillover effects influence our findings, we perform four robustness checks:

1) We conduct a closer examination of our control websites based upon their share of EU-traffic (using a quantile analysis as well as a linear regression).

2) We conduct a closer comparison of EU-websites and Non-EU-websites for Non-EU-users.

3) We conduct a closer comparison of EU-users and Non-EU-users for Non-EU-websites.

4) We examine the Non-EU-user interface on a subsample of control websites.

Another potential cause of concern might be that websites based in the EU could have shifted their geographic location. For example, shifting the location from an EU location to a Non-EU location would result in a website strategically evading the need for GDPR compliance for Non-EU-users. Such strategic shifts would again harm the validity of our control group as some websites would self-select themselves into the control group.

To examine whether potential strategic shifts performed by websites influence our findings, we conduct the following robustness check: We re-calculate the results of the GDPR's effect on websites for our main metric for a subsample of websites using a stricter control group that includes only websites with Non-EU-based domain suffixes.

*Closer Examination of the Non-EU-websites with no EU-traffic*

Although GDPR does not apply to Non-EU-websites when catering to Non-EU-users, websites might find it too costly to treat EU- and Non-EU-users differently and decide to comply with GDPR for Non-EU-users voluntarily. This voluntary compliance would represent a spillover effect of GDPR. From a logical perspective, such voluntary compliance is more likely for Non-EU-websites with a high share of EU-users. In contrast, if very few EU-users visit a Non-EU-website, that website likely has a much lower incentive to comply with GDPR for all users voluntarily. Thus, a spillover effect is more likely to exist for Non-EU-websites with a higher EU-user share.

In the first robustness check, we divide our control websites into deciles based on their pre-treatment EU-user traffic share for our main metric, the total number of visits. We can then examine whether a spillover effect exists for our control group by comparing the average number of visits before and after GDPR for each decile. As Table W 4 outlines, the number of visits from Non-EU-users to Non-EU-websites (the control group) increases after GDPR for all deciles, i.e., irrespectively of the EU traffic share. Thus, we find no cause for concern for a potential spillover effect within our control group.

*Table W 4: Examination of Control Group based on EU Traffic Share*

| Decile | | No. of Observations | EU-Share | Average Non-EU-Traffic | | |
|---|---|---|---|---|---|---|
| | | | | Pre-GDPR | Post-GDPR | Difference |
| No EU-traffic observed | Decile 0 | 594 | 0% | 137,992,462 | 250,139,478 | **+112,147,017** |
| EU-traffic observed | Decile 1 | 111 | 0.08% - 7.06% | 350,064,747 | 713,672,061 | **+363,607,314** |
| | Decile 2 | 111 | 7.11% - 15.82% | 297,563,325 | 587,272,620 | **+289,709,295** |
| | Decile 3 | 111 | 16.11% - 26.49% | 157,924,186 | 310,701,515 | **+152,777,330** |
| | Decile 4 | 111 | 26.60% - 52.58% | 41,281,851 | 83,942,652 | **+42,660,801** |
| | Decile 5 | 111 | 53.91% - 89.71% | 21,881,908 | 41,162,307 | **+19,280,399** |
| | Decile 6 | 111 | 90.16% - 98.49% | 2,552,483 | 5,300,178 | **+2,747,695** |
| | Decile 7 | 111 | 98.50% - 99.25% | 746,537.4 | 1,306,281.9 | **+559,744.5** |
| | Decile 8 | 111 | 99.26% - 99.55% | 404,036.1 | 803,807.5 | **+399,771.4** |
| | Decile 9 | 111 | 99.56% - 99.74% | 253,280.1 | 554,742.7 | **+301,462.6** |
| | Decile 10 | 108 | 99.74% - 99.95% | 163,518.2 | 389,103.4 | **+225,585.3** |

We further perform the following linear regression:

$$\log(1 + Total\ Visits_{Post-GDPR})$$
$$= \beta_0 + \beta_1 * \log(1 + Total\ Visits_{Pre-GDPR}) + \beta_2 * EU\ Share_{Pre\ GDPR} + \epsilon$$

Table W 5 displays the result: While $\beta_1$ is significant at the 0.01%-level, $\beta_2$ is not significant at the 5%-level. Thus, overall, the robustness check indicates that GDPR does not affect the results in a large enough way to pose a concern for a substantial change in effect sizes, and most certainly not in a potential change in the effect direction of our findings.

*Table W 5: Coefficients of Regression Analysis based on Control Websites' EU-Share*

|  | **Estimate** | **p-value** |
|---|---|---|
| **Intercept** | 0.928 *** (0.101) | 0.000 |
| **Log(1+Total Visits_{Pre-GDPR})** | 0.980 *** (0.006) | 0.000 |
| **EU-Share** | 0.044 (0.037) | 0.229 |

Notes: Standard errors are reported in parentheses.
Significance level: *** 0.1%-level ** 1%-level * 5%-level

*Closer Comparison of EU-websites and Non-EU-websites for Non-EU-users*

In the second robustness check, we examine the average traffic development of Non-EU-users on EU-websites (part of the treatment group) compared to Non-EU-websites (our control group). This examination can provide insights into whether there are spillover effects present for our control group. For the traffic examination, we focus on our main metric, the total number of visits. The mean comparison shown in Table W 6 shows that Non-EU-users visit EU websites (GDPR applies) less after GDPR. At the same time, the Non-EU-users visit Non-EU websites more after GDPR (GDPR does not apply). For those users, GDPR could only have an effect via a spillover effect.

*Table W 6: Closer Comparison of EU- and Non-EU-Websites' Total Visits for Non-EU-Users*

| **Website Location** | **User Location** | **Pre-/Post-GDPR** | **Average weekly visits** | **Difference** | **Difference-in-Difference** |
|---|---|---|---|---|---|
| EU-websites | Non-EU-users | Pre-GDPR | 3,159,196 | -321,937.4 | -555,845 |
| | | Post-GDPR | 2,837,259 | | |
| Non-EU-websites | | Pre-GDPR | 2,190,523 | +233,907.6 | |
| | | Post-GDPR | 2,424,430 | | |

*Closer Comparison of EU-users and Non-EU-users for EU-websites*

The third robustness check follows a similar approach and compares the traffic that Non-EU-websites get from EU-users (part of the treatment group) compared to Non-EU-users (control group). As shown in Table W 7, Non-EU-users visit Non-EU websites (GDPR does not apply) more after GDPR, while the EU-user base visit Non-EU websites less after GDPR (GDPR applies).

*Table W 7: Closer Comparison of EU- and Non-EU-Users' Total Visits for Non-EU-Websites*

| Website Location | User Location | Pre-/Post-GDPR | Average weekly visits | Difference | Difference-in-Difference |
|---|---|---|---|---|---|
| Non-EU-websites | EU-users | Pre-GDPR | 1,578,877 | -11,328.5 | -245,236.1 |
| | | Post-GDPR | 1,567,548 | | |
| | Non-EU-users | Pre-GDPR | 2,190,523 | +233,907.6 | |
| | | Post-GDPR | 2,424,430 | | |

The two robustness checks together show that both EU-users and Non-EU-users visit EU-websites less after GDPR. For EU-websites, GDPR applies to both user groups. Thus, both robustness checks show the same observation: For the website-instances for which GDPR applies, the traffic decreases after GDPR. For the website-instance for which GDPR does not apply, i.e., would only apply if websites voluntarily decided to comply (= spillover effect), the traffic increases after GDPR. Due to the contrarian development of our treatment group and our control group, the two robustness checks together show that it is unlikely that GDPR's effect spilled over to Non-EU-users on Non-EU-websites, at least on average. Overall, the three robustness checks indicate that there is no cause for concern for the adequacy of our control group and a potential influence of a spillover effect on our findings.

*Examination of Non-EU-User Interface on Subsample of Control Websites*

As a last robustness check to examine the potential spillover effect of GDPR to website-instances in the control group, we manually examine the user interface that a random sample of control websites displays to Non-EU-users. More specifically, we use the publicly available Internet Archive's Wayback Machine for 5% of the control websites, i.e., 85 control websites (https://web.archive.org/). Using the Wayback Machine, we accessed the past versions of websites via a crawler located in the US. Thus, these past versions represent the displayed content to Non-EU-users.

We accessed the past version of 69 control websites on the last day of our observation period, i.e., November 30[th], 2019. For the remaining 16 control websites, that date's version is not available. Thus, we accessed those websites on the first available date before November 30[th], 2019. We further accessed all websites on GDPR's enforcement date, i.e., May 25[th], 2018.

We observe that 94.12% of the control websites did not adjust their user interface in terms of the level of privacy after GDPR. Only 5.88% of the control websites slightly adjusted their user interface regarding the level of privacy between the GDPR's enforcement date and the end of our observation period. However, the observed changes do not result in the websites being compliant with GDPR. Thus, we observe no voluntary compliance to GDPR on the subsample of control websites within our observation period.

*Usage of Stricter Control Group with Websites with Definitive Non-EU-Location*

The last robustness check aims to account for a potential effect of possible strategic shifts performed by websites in their location. For instance, websites based in the EU before GDPR might have relocated their data processing location to a Non-EU location. Such a relocation would result in websites not having to comply with GDPR for Non-EU-users. While such relocations are not publicly known or observable, we examine the potential effect of such relocations.

More specifically, we construct a stricter control group and examine whether the observed effects of GDPR change significantly for our main metric. We construct the stricter control group by investigating the domain suffixes of websites (e.g., ".de," ".com"). We divide these domain suffixes for our original control websites according to whether the domain suffix 1) indicates an EU-website, 2) indicates a Non-EU-website, or 3) does not indicate a certain location. For the stricter control group, we only include the control websites of our original sample with domain suffixes that indicate a Non-EU location. The intuition behind this stricter control group is as follows: Websites with a domain suffix indicating a Non-EU location have a high probability of being a website not based in the EU, whereas websites with an EU-domain suffix have a higher probability of actually being a EU-based website. The latter had an incentive to shift their location to a Non-EU one. Thus, including only Non-EU-based suffixes ensures that the stricter control group includes only websites with a very low probability of having an EU-location before GDPR.

In our robustness check, we perform the same calculations as for our original control group. We then calculate the GDPR's short- and long-term effect on our main metric for a subsample of 226 websites. We observe that there is no significant difference between the GDPR's effect across the websites for our main metric for the short-term (average effect for the stricter control group for the subsample: -4.56%; average effect for the original control group for the subsample: -7.64%; no significant difference on 5%-level) or long-term (average effect for the stricter control group for the subsample: -19.90%; average effect for the original control group for the subsample: -18.64%; no significant difference on 5%-level) effects, irrespectively of whether we use our original or stricter control group. Thus, there is no cause for concern that strategic shifts might influence the validity of our control group. Furthermore, strategic shifts known to the public only correspond with a voluntary consideration of the GDPR requirements regarding data storage and data security – and do not directly affect the user interface, e.g., displaying a cookie banner or adjusting the privacy policy.

*WEB APPENDIX F: ROBUSTNESS CHECKS WITH RESPECT TO DATA*

We use a dataset provided by SimilarWeb. Although companies (e.g., Google, Alibaba) and researchers in top-tier academic journals (e.g., Calzada and Gill 2020, Lu et al. 2020) use SimilarWeb's data, it is not entirely clear how SimilarWeb collects its data and whether GDPR affects SimilarWeb in its data collection methods. For instance, if GDPR had affected SimilarWeb in its data collection methods post-GDPR, the validity of the data source, and thus our results, would be reduced.

Accordingly, we examine the quality and validity of SimilarWeb's data post-GDPR in two robustness checks. More specifically, we compare the quality of our data source SimilarWeb for a subset of websites with another highly reliable data source (German AGOF) for the:

1) number of unique visitors and

2) number of page impressions.

AGOF (https://www.agof.de/en/) collects high-quality and certified data on German websites, is widely used for media planning purposes, and is very transparent in its data collection procedure. AGOF can thus be considered the official gold-standard web traffic measurement in Germany. AGOF states that although GDPR does not directly impact its metrics, GDPR might affect its metrics due to users changing their interaction with the website due to changes that the website performed due to GDPR – which we aim to capture in our study.

*Comparison of SimilarWeb Data with AGOF Data for Unique Visitors*

We compare AGOF's reported data with our available data from SimilarWeb for the metric for the websites, period, and user base that overlap across the two datasets: The number of unique visitors for 23 websites for 2018. We perform a linear regression between the two data sources of the following form:

$$\log(1 + SimilarWeb) = \beta_0 + \beta_1 * \log(1 + AGOF) + \beta_2 * Postperiod + \epsilon$$

We find the following results shown in Table W 8: While $\beta_1$ is significant at the 0.01%-level, $\beta_2$ is not significant at the 5%-level. Thus, overall, the robustness check indicates that GDPR does not affect the results in a large enough way to pose a concern for a substantial change in effect sizes, and most certainly not in a potential change in the effect direction of our findings.

*Table W 8: Results of Regression Analysis comparing the Number of Unique Visitors for SimilarWeb and AGOF*

|  | Estimate | p-value |
|---|---|---|
| **Intercept** | 4.444 *** (0.124) | 0.000 |
| **Log(1+AGOF)** | 0.755 *** (0.009) | 0.000 |
| **Postperiod** | 0.039 (0.022) | 0.082 |
| **Number of observations** | 23 | 23 |
| **R²** | 0.8697 | |
| **Adj. R²** | 0.8695 | |

Notes: Standard errors are reported in parentheses.
Significance level: *** 0.1%-level ** 1%-level * 5%-level

*Comparison of SimilarWeb Data with AGOF Data for Page Impressions*

We compare AGOF's reported data with our available data for the metric for the websites, period, and user base that overlap across the two datasets: The number of page impressions for 23 websites for 2018. We perform a linear regression between the two data sources of the following form:

$$\log(1 + SimilarWeb) = \beta_0 + \beta_1 * \log(1 + AGOF) + \beta_2 * Postperiod + \epsilon$$

We find the following results shown in Table W 9: While $\beta_1$ is significant at the 0.01%-level, $\beta_2$ is not significant at the 5%-level. Thus, overall, the robustness check indicates that GDPR does not affect the results in a large enough way to pose a concern for a substantial change in effect sizes, and most certainly not in a potential change in the effect direction of our findings.

*Table W 9: Results of Regression Analysis comparing the Number of Page Impressions for SimilarWeb and AGOF*

|  | **Estimate** | **p-value** |
|---|---|---|
| **Intercept** | 4.133 *** (0.098) | 0.000 |
| **Log(1+AGOF)** | 0.729 *** (0.006) | 0.000 |
| **Postperiod** | -0.013 (0.022) | 0.555 |
| **Number of observations** | 23 | 23 |
| **R²** | 0.9304 | |
| **Adj. R²** | 0.9303 | |

Notes: Standard errors are reported in parentheses.
Significance level: *** 0.1%-level ** 1%-level * 5%-level

For our calculation of GDPR's effect on websites, no other changes in potentially confounding factors should coincide with the enforcement of GDPR. Substantial changes in confounding factors other than GDPR might have affected only the traffic of Non-EU-users on Non-EU-websites and have not affected EU-users or Non-EU-users that visit EU-websites. Such changes could influence the validity of our findings. Therefore, we investigate whether there have been substantial changes in factors other than the regulatory framework (e.g., internet speed) for Non-EU-users and EU-users.

More specifically, we examine any observable changes between the control website-instances and our treated website-instances other than the GDPR. We examine the internet speed, the share of people with access to the internet, the share of people using laptops, and the share of people using smartphones across our user base. These four potentially confounding factors might affect users' browsing behavior because a higher speed might result in browsing more websites in a shorter period. A higher share of the population with access to the internet, laptops, or smartphones can affect the number of online users.

While it is challenging to examine changes on the website-instance level, we examine changes on the user location level: As a proxy for the EU-user base, we examine German users and as a proxy for the Non-EU-user base, we examine US users. Although this separation does not account that GDPR affects Non-EU-users if they visit an EU-website, it can proxy whether there were changes between the two user groups.

As shown in Table W 10, we do not observe substantial differences between Non-EU-users and EU-users in the pre- and post-GDPR comparison for the selected confounding factors. If anything, the selected confounding factors indicate that EU-users should have exhibited an increased browsing behavior compared to the US group. Thus, this robustness check indicates that the above-mentioned other parameters did not negatively affect the user behavior and did not undermine the validity of our findings.

*Table W 10: User Group Comparison of Confounding Factors*

| Confounding Factor | User Group | Before GDPR: January 2018 | After GDPR: January 2019 |
|---|---|---|---|
| **Average Mobile Internet Connection Speed** | EU (Germany) | 26.43 Mbps | 31.69 Mbps *(+20% YoY-Growth)* |
| | Non-EU (USA) | 27.22 Mbps | 32.01 Mbps *(+18% YoY-Growth)* |
| **Penetration of Internet among Population** | EU (Germany) | 91% | 96% *(+5.49% YoY-Growth)* |
| | Non-EU (USA) | 88% | 95% *(+7.95% YoY-Growth)* |
| **Share of Population Using Laptops or Desktop** | EU (Germany) | 76% | 76% *(+0.00% YoY-Growth)* |
| | Non-EU (USA) | 77% | 77% *(+0.00% YoY-Growth)* |
| **Share of Population Using Smartphones** | EU (Germany) | 75% | 75% *(+0.00% YoY-Growth)* |
| | Non-EU (USA) | 78% | 78% *(+0.00% YoY-Growth)* |

*Source: Datareportal Reports – Digital 2018 and Digital 2019*

*WEB APPENDIX H: ROBUSTNESS CHECKS*
*WITH RESPECT TO THE SYNTHETIC CONTROL METHOD*

When calculating GDPR's effects in our analysis, we had to make several decisions: the method used, the requirement of the control and treated websites being in the same industry, and the number of control websites. As these decisions might impact our results, we examine the sensitivity of our estimates to the selected specifications.

To calculate GDPR's effect on websites, we used the SCG method and applied it using the R package "gsynth." However, apart from the specific method used in the "gsynth" package, other packages and corresponding methods might result in different findings. Furthermore, within our chosen SCG method, we had to make several decisions and assumptions to select the control websites. More specifically, we limited the number of control websites to five to avoid overfitting and required the control websites to belong to the same industry as our treated website. If the results are sensitive to our chosen method and the different decisions, these two decisions might affect our findings and thus the validity of our results. We thus examine whether the usage of our chosen methodology and our decisions influence our findings. We perform four robustness checks:

1) We re-calculate the results of the GDPR's effect on websites for our main metric using a different package for the SCG method: the "synth" package corresponding to the original SCG method (e.g., Abadie et al. 2015).

2) We calculate the GDPR's effect on websites for our main metric for a subsample of the websites without requiring the control websites to be in the same industry as the treated website, i.e., only selecting the control websites based on their correlation with the respective treated website.

3) We re-calculate the results of the GDPR's effect on websites for our main metric for a subsample of the websites requiring the control websites to have the same or a similar share of EU-traffic as the treated website. Thus, the treated and control websites would have similar incentives to be GDPR-compliant, as discussed in Web Appendix D.

4) We re-calculate the results of the GDPR's effect on websites for our main metric for a subsample of the websites with ten instead of five control websites for the synthetic control group.

*Usage of the Another Synthetic Control Method*

Our analysis applies the generalized SCG method corresponding to the "gsynth" package in R, a popular R package and method for SCG. In this first robustness check, we examine whether using another original SCG Method results in similar values of the observed effects. More specifically, we compare the results obtained when using the "gsynth" package to another popular SCG package: the "synth" package. We calculate the GDPR's short- and long-term effects on our main metric for a subsample of 3,455 website-instances. We find no significant difference between the results of the "gsynth" and "synth" packages for the short-term effects (average effect for "synth": -4.96%; average effect for "gsynth": -6.88%; no significant difference on 5%-level) or long-term effects (average effect for "synth": -18.39%; average effect for "gsynth": -19.29%; no significant difference on 5%-level).

*Calculation of Synthetic Control Group without Industry Specification*

We examine whether requiring the control websites of the SCG to be in the same industry as the treated website impacts our findings. Thus, we remove the industry specification and calculate the correlation of all possible control website-instances with the treated-website instance. The SCG thus includes the five control website-instances with the highest correlation with the treated website-instance irrespectively of the websites' industry. We then calculate the GDPR's short- and long-term effects for a subset of 100 treated website-instances. The robustness check shows that not requiring the control website-instances to be in the same industry as the treated website-instance results in no significant difference in the findings for the short-term effects (average effect for new specification: -2.84%; average effect for original specification: -6.69%; no significant difference on 5%-level) or long-term effects (average effect for new specification: -5.65%; average effect for original specification: -8.50; no significant difference on 5%-level).

*Calculation of Synthetic Control Group with EU-Traffic Share as Matching Variable*

As we discussed in the prior section in which we examined the validity of our control group, websites with a similar share of EU-traffic might have a similar compliance incentive. Thus, to account for this potential incentive-alignment with the EU-traffic share, we conduct the following robustness check: In our selection of the five control website-instances for the SCG calculation, instead of limiting the potential pool of control website-instances to the ones within the same industry as the treated website-instance, we require the potential control website-instances to have the same or a similar EU-share as the treated website-instance. We then re-calculate the GDPR's short- and long-term effects for our main metric for a subsample of 393 websites.

The robustness check shows that not requiring the control website-instances to be in the same industry but to have the same EU-share as the treated website-instance results in no significant difference in the findings for the short-term effects (average effect for new specification: -7.07%; average effect for original specification: -3.79%; no significant difference on 5%-level) or long-term effects (average effect for new specification: -4.21%; average effect for original specification: -5.77%; no significant difference at 5%-level).

*Calculation of Synthetic Control Group with a Higher Number of Control Websites*

We limit the number of control website-instances for the SCG calculation to the five website-instances that have the highest pre-treatment correlation with our treated website-instance to avoid overfitting. This robustness check examines whether our estimates are robust to an increase in the number of control website-instances selected for the SCG calculation. Thus, in selecting the control website-instances for the SCG calculation, we select the ten instead of five control website-instances with the highest pre-treatment correlation for a subsample of 60 treated website-instances. We then re-calculate the GDPR's short- and long-term effects for our main metric for the subsample. The robustness check shows that our presented findings in this study are robust to an increase in the number of control website-instances to ten instead of five, i.e., there is no significant difference in the findings for the short-term (average effect for 10 control websites: -13.91%; average effect for 5 control websites: -12.62%; no significant difference on 5%-level) or long-term effects (average effect for 10 control websites: -17.36%; average effect for 5 control websites: -14.87%; no significant difference on 5%-level).